\documentclass[reqno,12pt]{amsart}

\usepackage[centertags]{amsmath}
\usepackage{amsfonts}
\usepackage{amssymb}
\usepackage{amsthm}
\usepackage{newlfont}

\theoremstyle{plain}
  \newtheorem{theorem}{Theorem}[section]
  
  \newtheorem{proposition}[theorem]{Proposition}
  
\theoremstyle{definition}

\theoremstyle{remark}

\numberwithin{equation}{section}

 \let\be=\beta \let\de=\delta 
\let\ve=\varepsilon   
 \let\la=\lambda \let\om=\omega 
\let\si=\sigma

\let\De=\Delta  \let\La=\Lambda \let\Om=\Omega
\let\Th=\Theta

\newcommand{\caE}{{\mathcal E}}

\newcommand{\caO}{{\mathcal O}}

\newcommand{\bbE}{{\mathbb E}}

\newcommand{\bbR}{{\mathbb R}}

\newcommand{\bbZ}{{\mathbb Z}}

\newcommand{\opunit}{\text{1}\kern-0.22em\text{l}}

\newcommand{\bsE}{{\boldsymbol E}}

\newcommand{\bsP}{{\boldsymbol P}}

\newcommand{\ie}{i.e.\;}

\newcommand{\rel}{\,|\,}

\DeclareMathOperator{\prob}{Prob}

\begin{document}

\begin{center}
\noindent{\large \bf TIME-REVERSAL AND ENTROPY} \\

\vspace{15pt}

{\bf Christian Maes}\footnote{email: {\tt
christian.maes@fys.kuleuven.ac.be}}
and {\bf Karel Neto\v cn\'y}\\
Instituut voor Theoretische Fysica\\ K.U.Leuven, Belgium.

\end{center}

\vspace{20pt} \footnotesize \noindent {\bf Abstract: } There is a
relation between the irreversibility of thermodynamic processes as
expressed by the breaking of time-reversal symmetry, and the
entropy production in such processes. We explain on an elementary
mathematical level the relations between entropy production,
phase-space contraction and time-reversal starting from a
deterministic dynamics. Both closed and open systems, in the
transient and in the steady regime, are considered.  The main
result identifies under general conditions the statistical
mechanical entropy production as the source term of time-reversal
breaking in the path space measure for the evolution of reduced
variables.  This provides a general algorithm for computing the
entropy production and to understand in a unified way a number of
useful (in)equalities.  We also discuss the Markov approximation.
Important are a number of old theoretical ideas for connecting the
microscopic dynamics with thermodynamic behavior.

\vspace{20pt} \normalsize

\section{Introduction}\label{sec: introduction}

An essential characteristic of irreversible thermodynamic
processes is that the time-reversal invariance of the microscopic
dynamics is apparently broken.  This means that out of equilibrium
a particular sequence of macrostates and its time-reversal can
have a very different plausibility.  This, basically, must be the
reason for the positivity of transport coefficients, or, more
generally, for the positivity of entropy production. It has
already been argued before in \cite{M,MRV,mm}, mostly via
examples, how there is a direct relation between entropy
production and the ratio of probabilities for time-reversed
trajectories.  Most of this was however concentrated on finding a
unifying framework for equalities and inequalities that have
recently appeared in nonequilibrium statistical mechanics,
generalizing, so it is hoped, close to equilibrium relations. Most
prominent among those is the symmetry expressed in the
Gallavotti-Cohen fluctuation theorem, \cite{ecm,GC}.  In the
present paper, we turn to more fundamental issues for identifying
the statistical mechanical definition of entropy production rate
and to offer a possible answer for various interpretational
problems that have remained. The emphasis is on the simplicity of
the explanation avoiding technical issues.


\section{Results}

Nonequilibrium statistical mechanics is to a large extent still
under construction.  Recently, there have been made various
proposals for a definition of statistical mechanical entropy
production going beyond the close to equilibrium regime and
through which fluctuations in irreversible processes could be
studied. In some cases, the theory of dynamical systems has been a
source of inspiration and it was argued that phase space
contraction can be identified with entropy production with
nonequilibrium ensembles obtained as limits of ergodic averages,
see e.g. \cite{GC, ru}.  A somewhat different approach started
from taking advantage of the Gibbsian structure of the
distribution of space-time histories where entropy production
appeared as the source term for time-reversal breaking in the
space-time action functional.  These two approaches have in fact
much in common, at least concerning the mathematical analysis, see
\cite{M}.  Since then, many examples have passed the test of
verifying that the various algorithms indeed give rise to the
physical entropy production.  There has however not been a
derivation from first principles to convince also the stubborn
that the algorithm of \cite{M,MRV} applied to models in
nonequilibrium dynamics to identify the entropy production, is
entirely trustworthy. The main result of the present paper is to
give such a derivation: that indeed under very general conditions,
both for closed systems and for open systems, both in the
transient regime and in the steady state regime, the entropy
production can be obtained as the source term of time-reversal
breaking in the action functional of the path space measure that
gives the distribution of the histories (on some thermodynmaic
scale) of the system.  This representation is useful because it
gives the entropy production as a function of the trajectories and
it allows easy mathematical manipulations for taking the average
(to prove that it is positive) and for studying the fluctuations
(to understand symmetries under time-reversal).

This paper is more or less self-contained with a first Section
\ref{sec: setup} introducing the main actors.  Sections \ref{sec:
er} and \ref{sec: cou} contain the main result. The difference is
that \ref{sec: er} is entirely about the transient regime for
closed systems, while Section \ref{sec: cou} deals with open
systems and discusses the steady state regime.  Sections
\ref{sec:markov} and \ref{sec: om} discuss their consequences in
the Markov approximation. Section \ref{sec: PSC} relates the
approach to results inspired by the theory of chaotic dynamical
systems, in particular how phase space contraction can play the
role of entropy production.  Along the way, we suggest
interpretations that we think are helpful for starting
nonequilibrium statistical mechanics.

\section{Set-up}\label{sec: setup}

\subsection{Phase space and microscopic dynamics}
Let $\Omega$ be the phase space of a closed isolated mechanical
system with $x \in \Omega$ representing the microstates, i.e., as
described by canonical variables for a classical system of $N$
particles, $x = (q_1,\ldots,q_N,p_1,\ldots,q_N)$.
The Hamiltonian dynamics specifies the flow $x\mapsto \phi_t(x)$
on $\Omega$ under which $x$ (at some initial time $t_0$) evolves
into $\phi_t(x)$ at time $t_0+t$. The dynamics is  reversible in
the sense that $\pi\phi_t\pi=\phi_t^{-1}$ where the time-reversal
$\pi$ on $\Omega$ is the involution that changes the sign of the
momenta $p_i$. The flow preserves the phase space volume
(Liouville's theorem); the Jacobian determinant equals unity,
$|d\phi_t(x)/dx|=1$
for each $t$ and the Liouville measure $dx$ is time-invariant. \\
We fix a time-interval $\delta$ and write $f\equiv \phi_{\delta}$.
Of course, $f$ preserves the phase space volume and $\pi f \pi =
f^{-1}$.

\subsection{Reduced variables}

The time evolution preserves the total energy. We introduce
therefore the state space $\Omega_E\equiv \Gamma$, the energy
shell, corresponding to a fixed total energy $E$ or better, some
interval around it. We denote by $|A|$ the phase space volume of a
region $A\subset \Gamma$ given by the projection $\rho$ of the
Liouville measure into $\Gamma$.  Since $\Gamma$ is thought of as
containing a
 huge number of degrees of freedom, it is reasonable to divide it further.
For comparison with experimental situations, we look at some
special set of variables, suitably defined in each case, which
give a more coarse-grained, contracted, or reduced description of
the system, \cite{om,Z,le}. Depending on the context or on the
application, their precise nature may vary.  It could be that we
look at macroscopic variables $\alpha(x)$ implying a subdivision
of $\Gamma$ by cutting it up in phase cells defined by $a <
\alpha(x) < a + \Delta a$ (with some tolerance $\Delta a$), or
that we split up $x=(y,z)\in \Gamma$ into an observable part $y$
and a background part $z$. For example, the $y$ might refer to the
coordinates of the particles in a subsystem while the background
is only monitored as the macrostate of reservoir(s).\\ At this
moment, we do not commit ourselves to one picture but rather
imagine somewhat abstractly a map $M: \Gamma \rightarrow
\hat\Gamma: x\mapsto M(x)$ where $\hat\Gamma$ is the reduced phase
space, a finite partition of $\Gamma$.  When having in mind
macrostates, this space $\hat\Gamma$ would correspond to the
$\mu-$space of Gibbs. The fact that this partition is assumed
finite is not realistic, it is more like $\bbR^d$, but it is
convenient for the notation and it is not essential.  With some
abuse of notation, the elements of $\hat\Gamma$ are denoted by $M$
(standing for all possible values of $M(x)$) and of course, every
microscopic trajectory $\gamma = (x,fx,\ldots,f^nx)$ gives rise to
a trajectory $\omega = (M(x),M(fx),\ldots,M(f^nx))$ in
$\hat\Gamma$.
We also assume for simplicity that $\pi M$ is well defined via
$\pi M = M\pi$, that is $Mx=My$ implies $M\pi x=M\pi y$, for all $x,y \in \Gamma$.\\

\subsection{Distributions}\label{dis}
Probabilities enter the description because the exact microstate
of the system is not accessible to us.  This is so when preparing
the system and also later when we observe the system.  Even when
we know the reduced state, we still need to evaluate the
plausibility of background configurations in order to predict the
future development on the level of the reduced states.  A natural
choice here is to use the microcanonical ensemble.  That is, we
sample the reduced variables according to some probability
distribution $\hat\nu$ on $\hat\Gamma$ and we impose the
microcanonical distribution on each phase cell $M$. If $\hat\nu$
is a probability on $\hat\Gamma$, then
$\hat\nu\times\rho\,(x)\equiv \hat\nu(M(x))/ |M(x)|$ is the
probability density on $\Gamma$ obtained from $\hat\nu$ by uniform
randomization (microcanonical ensemble) inside each
$M\in\hat\Gamma$.  It is uniquely determined from the two
conditions (1) $\hat\nu\times\rho\,(M) = \hat\nu(M)$ and (2)
$\hat\nu\times\rho\,(x|M) = 1/|M|, \; x\in M$, $M\in \hat\Gamma$.
(Remark: the writing $\hat\nu\times \rho$ has no meaning in itself
except that it is the notation we use for this probability
density.)  In words, the probability of a microstate $x$ is the
probability (under $\hat\nu$) of its corresponding reduced state
$Mx$ multiplied with the {\it a priori} probability (under the
Liouville measure) of $x$ given the reduced state $Mx$. So if we
take $\hat\nu=\delta(M - \cdot)$ concentrated on the reduced state
$M\in\hat\Gamma$, then $\hat\nu\times\rho$ is the initial
probability density corresponding to an experiment where the
system is started in equilibrium subject to constraints; that is a
uniform (i.e., microcanonical)
distribution of the phase points over the set $M$. \\
For the opposite direction, we note that every density $\nu$ on
$\Gamma$ gives rise to its projection $p(\nu)$, a probability on
$\hat\Gamma$, via
\[
p(\nu)(M) \equiv \nu(M) = \int dx \nu(x) \delta(M(x)-M)
\]
and obviously, $p(\hat\nu\times\rho)=\hat\nu$.  All this is very
much like what enters in projection-operator techniques,
\cite{Z}.\\
It now makes sense to ask for  the probabilities on $\Gamma$ at
time $t$, given that the system started at time zero in $M_0 \in
\hat\Gamma$; we always mean by this that the microstates were
uniformly sampled out of $M_0$. They are given by the ratio
\begin{equation}\label{ratio}
\mbox{Prob}[\phi_t(x) \in A \rel M(x) = M_0] \equiv
\frac{|\phi_t^{-1}A\cap M_0|}{|M_0|}
\end{equation}
More generally, when, for all we know at time zero, the statistics
of the reduced variables is given in terms of the probability
$\hat\nu$ on $\hat\Gamma$, then, at time $t$, the statistics on
$\hat\Gamma$ is obtained from
\[
\hat\nu_t \equiv p((\hat\nu\times\rho)_t)
\]
where $(\hat\nu\times\rho)_t$ gives the distribution at time $t$
as solution of the Liouville equation with initial distribution
$\hat\nu\times\rho$ on $\Gamma$.\\
Finally, given an initial probability $\hat\nu$ on $\hat\Gamma$,
we can look at the collection of all paths
$\omega=(M(x),M(fx),\ldots,M(f^nx)), (n$ fixed) where, first, a
reduced state $M_0$ is drawn according to $\hat\nu$ and then, with
uniform probability on $M_0$ a microstate $x\in M_0$ is drawn (so
that $M(x)=M_0$). We denote the resulting distribution on these
paths by $\bsP_{\hat\nu}$; it defines the path space measure on
trajectories in $\hat\Gamma$.

\subsection{Entropies} \label{suben}
There will be various types of entropies appearing in what
follows, each having their specific role and meaning.  There is
first the Shannon entropy $S(\mu)$, a functional on probability
densities $\mu$  on $\Gamma$:
\begin{equation}\label{shan}
S(\mu) \equiv - \int dx \mu(x) \ln \mu(x)
\end{equation}
We can also define the Shannon entropy $\hat{S}(\hat\mu)$ for
probability laws on $\hat\Gamma$ through
\begin{equation}\label{shan1}
\hat{S}(\hat\mu) \equiv - \sum_M \hat\mu(M) \ln \hat\mu(M)
\end{equation}
There is secondly the Boltzmann entropy $S_B$ which is first
defined on $M\in\hat\Gamma$, and then for each microstate $x\in
\Gamma$ as
\begin{equation}\label{boltz}
\hat{S}_B(M) \equiv \ln |M|; \quad S_B(x) \equiv \hat{S}_B(M(x))
\end{equation}
The dependence on the number of particles $N$ is ignored here as
it shall be of no concern. Most frequently, we have in mind here
macroscopic variables (such as density and/or velocity profile(s))
for characterizing the reduced states. Any two microstates on
$\Gamma$ are {\it a priori} equivalent but if we randomly pick a
microstate $x$ from $\Gamma$, the chance that its reduced state
$M(x)$ equals  $M\in \hat\Gamma$ increases with greater Boltzmann
entropy $\hat{S}_B(M)$.  We can then expect, both for the forward
evolution and for the backward evolution (positive or negative
times) that the Boltzmann entropy should increase.  This
time-reflection invariance of the increase of entropy is an
instance of the dynamic reversibility and it interprets
 the paradoxical words of Boltzmann when speaking about the increase of
 entropy (minus the $H-$ functional)
``that every point of the $H-$curve is a maximum,'' see \cite{K}.
From this, equilibrium is understood as the state of maximal
entropy, given the constraints in terms of macroscopic values that
define the equilibrium conditions (such as energy, volume and
number of particles). Upon varying the constraints, this maximal
Boltzmann entropy will behave as the thermodynamic entropy
(defined operationally).  Yet, even out of equilibrium the
Boltzmann entropy makes sense which is essential and needed even
to discuss fluctuations around equilibrium.  It then continues to
correspond to the thermodynamic entropy in the close to
equilibrium treatments of irreversible processes.\\
The Boltzmann entropy thus tells how typical a macroscopic
appearance is from counting its possible microscopic realizations.
Also the Shannon entropy has its origin in counting (for example
in evaluating Stirling's formula or other combinatorial
computations) and it is therefore not surprising that there are
relations between the two. For our context, the following identity
between Shannon and Boltzmann entropies holds:
\begin{equation}\label{shabol}
S(\hat\nu\times\rho) - \hat{S}(\hat\nu) = \sum_M \hat\nu(M)
\hat{S}_B(M)
\end{equation}
\\
Thirdly, we will need the Gibbs entropy $S_G(\hat\nu)$ which is a
functional on the statistics $\hat\nu$ of reduced states:
\begin{equation}\label{gibbs}
S_G(\hat\nu) \equiv \sup_{p(\mu)=\hat\nu} S(\mu)
\end{equation}
Equivalently,
\begin{equation}\label{gibbs1}
S_G(\hat\nu) = S(\hat\nu\times \rho)
\end{equation}
because we always have $p(\hat\nu\times\rho)=\hat\nu$ and a
standard computation shows that $S(\hat\nu\times\rho) \geq S(\mu)$
for every density $\mu$ on $\Gamma$ for which also
$p(\mu)=\hat\nu$ (Gibbs variational principle).
At the same time, from (\ref{shabol}), note that for
$\hat\nu=\delta(M-\cdot)$,
\[
\hat{S}_B(M) =  S(\hat\nu\times\rho)
\]
Combining this with (\ref{gibbs1}), we observe that in case
$\hat\nu$ concentrates on one $M\in\hat\Omega$, then the Boltzmann
and the Gibbs entropies coincide:
\begin{equation}\label{bolgi}
\hat{S}_B(M) = S_G(\delta(M-\cdot))
\end{equation}
which indicates that the Gibbs entropy is, mathematically
speaking, an extension of the Boltzmann entropy since it lifts the
definition of $\hat{S}_B$ to the level of distributions on
$\hat\Gamma$.

Another application of the Gibbs variational formula (\ref{gibbs})
concerns the change of entropy under the Hamiltonian dynamics.
This was emphasized by Jaynes, see e.g. \cite{j}.   Suppose the
system is initially (at time $t_0=0$) prepared with density
$\hat\nu\times \rho$ (i.e., microcanonically with reduced states
sampled from $\hat\nu$). Then, according to the Liouville
equation, at time $t$ we obtain the density
$(\hat\nu\times\rho)_t$.  But only the reduced state is monitored,
i.e., its projection $p((\hat\nu\times\rho)_t)=\hat\nu_t$.  This
is for example obtained from the empirical distribution of the
macrovariables. From (\ref{gibbs}) it follows that $S_G(\hat\nu_t)
\geq S_G(\hat\nu)$ because, by Liouville's theorem,
$S((\hat\nu\times\rho)_t)=S(\hat\nu\times\rho)=S_G(\hat\nu)$.  We
call the difference
\[
S_G(\hat\nu_t) - S_G(\hat\nu) = \mbox{the (Gibbs-) entropy
production}
\]
It is always non-negative.  From (\ref{bolgi}), if we initially
prepare the system in some specific reduced state
$M_0\in\hat\Gamma$, then this (Gibbs-) entropy production equals,
in fact, $S_G(\hat\nu_t) - S_B(x)\geq 0, x\in M_0$.  If the set of
reduced variables allows a hydrodynamic description in which,
reproducibly for almost all $x\in M_0$, $M(\phi_tx) = M_t\in
\hat\Gamma$, then the experimentalist will, for all practical
purposes, identify $\hat\nu_t$ with $\delta(M_t-\cdot)$ and the
(Gibbs-) entropy production is then given by the change
$\hat{S}_B(M_t)- \hat{S}_B(M_0) = S_B(\phi_tx) - S_B(x)$ in
Boltzmann entropy. In other words, the (Gibbs-) entropy production
then coincides with the \mbox{(Boltzmann-)} entropy production.
Yet, from this we see that, under second law conditions, the
inequality $S_G(\hat\nu_t) \geq S_G(\hat\nu)$ obtained from the
Gibbs variational principle is doing great injustice to the actual
difference between initial and final Boltzmann entropies: the
value of $\hat{S}_B(M_0)$ will be very small compared to the
equilibrium entropy $\ln |\Omega_E|$ ($\simeq \hat{S}_B(M_t)$ for
$|M_t| \simeq |\Gamma|$) if $M_0$ corresponds to a preparation in
a special nonequilibrium state.  As it is often correctly
emphasized, the stone-wall character of the second law derives
from the great discrepancy in microscopic and macroscopic scales
as a result of the huge number of degrees of freedom in a
thermodynamic system. Moreover, a theoretical advantage of
considering $S_B(\phi_tx)- S_B(x)$ is that this is directly
defined on the phase space $\Gamma$ and in fact allows a
microscopic derivation of the second law based on statistical
considerations concerning the initial state, see \cite{Gold,Lebo}
for recent discussions. Note however that this advantage  also
implies the challenge to relate the Boltzmann entropy with more
operational definitions of entropy as practiced in thermodynamics
of irreversible processes where entropy production appears as the
products of fluxes and forces, as obtained from entropy balance
equations.  Yet, irreversible thermodynamics is restricted by the
assumption of local equilibrium
whose validity requires systems close to equilibrium.\\
Finally, there is the dynamical entropy $S_K$ that is an immediate
extension of the Boltzmann entropy but defined on trajectories:
for a given trajectory $\omega = (M_0,M_1,\ldots,M_n)$ in
$\hat\Gamma$, we put
\begin{equation}\label{sk}
S_K(\omega) \equiv \ln |\cap_{j=0}^n f^{-j}M_j|
\end{equation}
counting the microstates $x\in \Gamma$ for which $M(f^jx) = M_j
(\equiv \omega_j)$, $j=0,\ldots,n$.   In ergodic theory, this
dynamical entropy (\ref{sk}) is related to the Kolmogorov-Sinai
entropy via Breiman's theorem, \cite{Sinai}. In this way, the
Kolmogorov-Sinai entropy gives the asymptotics of the number of
different types of trajectories as time tends to infinity. Note
however that in the physical case we have in mind, $\hat\Gamma$
does not correspond to some kind of mathematical coarse graining
and there is no way in which it is assumed generating nor do we
intend to let it shrink $\hat\Gamma\rightarrow \Gamma$.

\subsection{Transient versus steady state regime}\label{trareg}

The family of nonequilibrium states is much more rich and varied
than what is encountered in equilibrium. It is often instructive
to divide nonequilibrium phenomena according to their appearance
in the transient versus the steady state regime. The simplest
example of a transient regime is when the total system starts from
a nonequilibrium state and is allowed to relax to equilibrium.
Steady state on the other hand refers to a maintained
nonequilibrium state. For this we need an open system that is
driven away from equilibrium by an environment.  All of this
however strongly depends on the type of variables that are
considered and over what length and time scales. In this paper we
take the point of view that the steady state is a special or
limiting case of the transient situation. Since the fundamental
dynamics is Hamiltonian for a closed system, any attempt to give a
microscopic definition of entropy production must start there. We
can then discuss the limiting cases or approximations through
which we are able to define also the mean entropy production rate
in the steady state.   In any case, the identification of the
statistical mechanical entropy production as function of the
trajectory must be independent from the regime, be it transient or
steady.

\section{Entropy production: closed systems}\label{sec: er}

Our main goal in the following two sections is to show how, via
time-reversal, we can define a function on path space which will
be recognized as the variable or statistical mechanical entropy
production.  It enables us to compute the time-derivative of the
entropy production, the entropy production rate.  Since this
function is defined on trajectories, we can also study its
fluctuations.\\
The situation to have in mind here is that of the transient regime
in a closed system; entropy production is the change of entropy.

Recall (\ref{ratio}). We start by observing that for any two
reduced states $M_0,M_n\in \hat\Gamma$ and for every microscopic
trajectory $\gamma$ corresponding to a sequence of microstates
starting in $M_0$ and ending in $M_n$,
\begin{multline}\label{microob}
  \ln \frac{\prob[(x,fx,\ldots,f^nx)= \gamma \rel M(x) =M_0]}
  {\prob[(x,fx,\ldots,f^nx)= \gamma\Theta \rel M(x) = \pi M_n]} =
\\
  \hat{S}_B(M_n) - \hat{S}_B(M_0)
\end{multline}
where $\gamma\Theta$ is the time-reversed microscopic trajectory.
That $\gamma\Theta$ is also a microscopic trajectory is an
expression of dynamic reversibility. This identity (\ref{microob})
follows because, given the initial reduced state, the probability
that a specific microscopic trajectory is realized only depends on
the probability of the initial microstate. But since we know to
what reduced state it belongs, that probability is just the
exponential of minus the Boltzmann entropy.

While the previous relation indicates that time-reversal
transformations are able to pick up the entropy production, we
cannot in practice sample microscopic trajectories.  In order to
lift these relations to the level of trajectories on $\hat\Gamma$,
we should also relax the condition that we start in a fixed
reduced state; if we only know the reduced state the dynamics
started from, we will not know in what specific reduced state we
land after time $t$.
For this additional uncertainty, there is a small price to be paid.\\
Let $\omega=(M_0,M_1,\ldots,M_n)$ be a possible trajectory on
$\hat\Gamma$. Its time-reversal is $\omega\Theta=(\pi
M_n,\ldots,\pi M_0)$. Let $\hat\mu$ and $\hat\nu$ be two
probabilities on $\hat\Gamma$. We ask for the ratio of
probabilities that the actual trajectory coincides with $\omega$
and with $\omega\Theta$, conditioned on starting the microscopic
trajectory sampled from $\hat\mu\times\rho$ and from
$\hat\nu\pi\times \rho$ respectively:
\begin{equation}\label{prop}
\frac{\prob_{\hat\mu\times\rho}[\mbox{trajectory }=\omega]}
{\prob_{\hat\nu\pi\times\rho}[\mbox{trajectory }=\omega\Theta]} =
\frac{\hat\mu(M_0)}{\hat\nu(M_n)}\frac{|M_n|}{|M_0|}
\end{equation}
More precisely, this wants to say that the corresponding path
space measures have a density with respect to each other, given by
\begin{equation}\label{prop1}
\frac{d\bsP_{\hat\mu}}{d\bsP_{\hat\nu\pi}\Theta}(\omega) =
\frac{\hat\mu(M_0)}{\hat\nu(M_n)}\frac{|M_n|}{|M_0|}
\end{equation}
To prove this, it suffices to see that, on the one hand
\[
\hat\mu\times\rho\,(\cap_{j=0}^nf^{-j}M_j)=
\frac{\hat\mu(M_0)}{|M_0|} |\cap_{j=0}^n f^{-j} M_j|
\]
and, on the other hand, for the denominator in the left-hand side
of (\ref{prop})-(\ref{prop1}),
\begin{equation}\nonumber
  |\cap_{j=0}^n f^{-j} \pi M_{n-j}| = |\cap_{j=0}^n f^{j} M_{n-j}|
  =  |\cap_{j=0}^n f^{j-n}  M_{n-j}|
\end{equation}
where we first used $\pi f^{-1} \pi = f$ and then the stationarity
of the Liouville measure under $f$.  Hence, the factor
$|\cap_{j=0}^n f^{-j} M_j|$ will cancel when taking the ratio as
in (\ref{prop})-(\ref{prop1}); this expresses the time-reversal
invariance of the dynamical entropy,
$S_K(\omega)=S_K(\Theta\omega)$, which excludes it as candidate
for entropy production.

The most interesting case is obtained by taking $\hat\nu
=\hat\mu_t$ for $t=n\delta$ in (\ref{prop}) or (\ref{prop1}).
Remember that $\hat\mu_t=p((\hat\mu\times \rho)_t)$ is the
projection on the reduced states of the measure at time $t$ when
started from the constrained equilibrium $\hat\mu\times \rho$. We
then get as a direct consequence of (\ref{prop1}):
\begin{proposition}\label{cor}
For every probability $\hat\mu$ on $\hat\Gamma$,
\begin{equation}\label{totalen}
\ln \frac{d\bsP_{\hat\mu}}{d\bsP_{\hat\mu_t\pi}\Theta}(\omega)=
[S_B(\phi_t x) - S_B(x)] +  [-\ln \hat\mu_t(M(\phi_t x)) + \ln
\hat\mu(Mx)]
\end{equation}
for all trajectories $\omega=(M(x),M(fx),\ldots,M(f^nx))$ in
$\hat\Gamma$, $x\in \Gamma$, $t=n\delta$.
\end{proposition}
The right-hand side of (\ref{totalen}) contains two contributions.
The first difference of Boltzmann entropies has already appeared
(alone) in (\ref{microob}) when the comparison was made between
probabilities for microscopic trajectories.  The second
contribution to (\ref{totalen}) can thus be viewed as originating
from the `stochasticity' of the reduced dynamics.  Note in
particular that even when $\hat\mu$ is concentrated on some $M\in
\hat\Gamma$, then $\hat\mu_t$ is still smeared out over various
possible reduced states. Yet, this second contribution can be
expected to be very small under second law conditions. After all,
if $\hat\mu(M)=\delta(M-M(x))$, then $\hat\mu(Mx)=1$.  And if
$M(f^nx)$ is large in the sense that `typically' almost all $z\in
\Gamma$ get into $M(f^nx)$ after a sufficient time $t=n\delta$,
then, we expect, $|M(\phi_tx)\cap \phi_t M(x)| \simeq |M(x)|$ so
that by (\ref{ratio}), also $\hat\mu_t(M(\phi_tx)) \simeq 1$ and
hence only the first Boltzmann contribution survives.  Of course,
for smaller systems, there is hardly a notion of what is typical
but we can see what to expect in general.

Let $\bsE_{\hat\mu}$ stand for the expectation with respect to the
path space measure $\bsP_{\hat\mu}$.
\begin{proposition}\label{propo2}
Denote (\ref{totalen}) by
\[
R_{\hat\mu}^t \equiv
\ln\frac{d\bsP_{\hat\mu}}{d\bsP_{\hat\mu_t\pi}\Theta}
\]
Then,
\begin{equation}\label{expr}
\bsE_{\hat\mu}[e^{-R_{\hat\mu}^t}]= 1
\end{equation}
In particular, its expectation equals the (Gibbs-) entropy
production:
\begin{equation}\label{gep}
\bsE_{\hat\mu}[R_{\hat\mu}^t]= S_G(\hat\mu_t) - S_G(\hat\mu) \geq 0
\end{equation}
\end{proposition}
\begin{proof}
The identity (\ref{expr}) is the normalization
\[
\bsE_{\hat\mu}[\frac{d\bsP_{\hat\mu_t\pi}\Theta}{d\bsP_{\hat\mu}}]=1
\]
The relation (\ref{gep}) follows from (\ref{shabol}) and the
definition (\ref{gibbs})-(\ref{gibbs1}) of Gibbs entropy from
inspecting the expectation of the right-hand side of
(\ref{totalen}). The non-negativity of the (Gibbs-) entropy
production was already explained under Section \ref{suben} but it
can also be obtained from applying the Jensen (convexity)
inequality to (\ref{expr}).
\end{proof}

\noindent{\bf Remark 1:} The above calculations have relied
heavily on the structure $\hat\mu\times \rho$ of the
distributions.  We will see in Section \ref{sec: cou} what happens
in the case where the distribution has the form $\bar\mu\times\nu$
where $\bar\mu$ will refer to the distribution of a subsystem and
the $\nu$ takes into account the macrostate of the enviroment that
further constraints the evolution.

\noindent{\bf Remark 2:} One may wonder about the physical
significance of the terms $[-\ln \hat\mu_t(M(\phi_t x)) + \ln
\hat\mu(Mx)]$ appearing in (\ref{totalen}).  There is no general
answer: they have {\it a priori} nothing to do with entropy
production but their addition can become physically significant to
the same extent that $\hat\mu$ and $\hat\mu_t$ are physically
motivated.  Nevertheless we continue to call the right-hand side
of
\[
[-\ln \hat\mu_t(\omega_n) + \ln \hat\mu(\omega_0)] +
[\hat{S}_B(\omega_n) - \hat{S}_B(\omega_0)] =
R_{\hat\mu}^t(\omega)
\]
 the
total variable (or statistical mechanical) entropy production
$R_{\hat\mu}^t$. It coincides with the (Boltzmann-) entropy
production under second law conditions and its expectation is the
(Gibbs-) entropy production. Even though $R_{\hat\mu}^t$ does not
quite coincide with the change of Boltzmann entropy, it has a
useful structure (as ratio of two probabilities) making the
studies of its fluctuations much easier, see for example
(\ref{expr}). The amazing point is that while
$\bsP_{\hat\mu}(\omega)$ and $\bsP_{\hat\mu_t\pi}(\Theta\omega)$
both depend on the entire path
$\omega=(\omega_0,\omega_1,\ldots,\omega_n)$, their ratio is a
state function, only depending on the initial and final states
$\omega_0$ and $\omega_n$.   We will use it throughout the
following.\\

\section{Entropy production: open systems} \label{sec: cou}

 As a general
remark, in the set-up of the present paper, nonequilibrium steady
states correspond in reality to a transient regime for the whole
system over timescales short enough for the macrostates of the
reservoirs not to have changed appreciably while long enough for
the internal subsystem to have reached a stationary condition.  We
start however again from the Hamiltonian dynamics of the total
system.

We consider the situation where the complete system consists of
one smaller and one larger subsystem. The latter represents an
environment to which the smaller subsystem is coupled and it can
have the form of one or more thermal reservoirs, for instance. For
short, the smaller subsystem will simply be called system.  The
total phase space is $\Om = \Om^0 \times \Om^1$, where the
superscripts $0$ and $1$ stand for the system and for the
environment, respectively. The dynamics of the total system is
Hamiltonian and we use the same notation as introduced in Section
\ref{sec: setup}; namely $f$ is the discretized flow and the
time-reversal $\pi$ on $\Om$ is assumed to have the form $\pi =
\pi^0 \otimes \pi^1$ (we will soon forget these superscripts). The
reduced picture is given by the partition $\hat\Om$ of $\Om$, with
the product structure $\hat\Om = \hat\Om^0 \times \hat\Om^1$. One
choice could be taking $\hat\Om^0 = \Om^0$  in case the system has
only a few microscopic degrees of freedom, and the elements of
$\hat\Om^1$ corresponding to fixed values of the energies of the
individual reservoirs. Preparing the system in the initial state
$\hat\mu^0$ and, independently,  preparing the environment in
$\hat\mu^1$, we construct the initial distribution on $\Om$, from
which the microstates are sampled, as $(\hat\mu^0 \otimes
\hat\mu^1) \times \rho$.  At this moment the microscopic dynamics,
conserving the total energy, takes over and the system gets
coupled with the environment. We then get, as used in the previous
section and as defined in Section \ref{dis}, a path space measure
$\bsP_{\hat\mu^0 \otimes \hat\mu^1}$ for the trajectories.

It is convenient to rephrase the above construction in the
following, more formal, way to make a connection with the scenario
of the previous section. We introduce yet another coarse-graining
in which only the system is observed, while the macrostate of the
environment is ignored. It is defined via the map $\bar{p}:\hat\Om
\mapsto \hat\Om^0$ which assigns to every $(M, E) \in \hat\Om$ its
first coordinate $M \in \hat\Om^0$.  That means that we actually
deal with two successive partitions: $\hat\Om$ is a partition of
the original phase space $\Om$ (involving both system and
environment) and $\bar\Om$ is taken as a further partition of
$\hat\Om$. $\bar\Om$ can be identified with $\hat\Om^0$ involving
only the system's degrees of freedom. The elements of $\bar\Omega$
are written as $M$ and those of $\hat\Om$ as $(M,E)$. To every
$(y,z)\in \Om =\Om^0\times\Om^1$ we thus associate an element
$My\in\bar\Om$
and an element $(My,Ez)\in \hat\Om$.\\
On $\bar\Om$ we put the distribution $\bar\mu\equiv \hat\mu^0$
which stands for the initial statistics of the small subsystem. On
$\hat\Om$ we put the distribution $\hat\mu$ for which we only ask
that $\hat\mu(M,E) = \hat\mu(M) \hat\mu^1(E)$ thus representing
the preparation of the environment. The first and crucial
observation we make is that
\begin{equation}\label{ur}
(\hat\mu^0 \otimes \hat\mu^1) \times \rho = \bar\mu \times
(\hat\mu \times \rho)
\end{equation}
The right-hand side is defined as the generalization of the
randomization we introduced in Section \ref{dis}: given a
distribution $\nu$ on $\Omega$ and a distribution $\bar\mu$ on
$\bar\Om$ we let
\[
\bar\mu \times \nu\; (y,z) \equiv
\bar\mu(My)\frac{\nu(y,z)}{\nu(My)}
\]
Take here $\nu = \hat\mu\times \rho$, then
$\nu(y,z)=\hat\mu(My,Ez)/|My||Ez|$ and $\nu(My) = \hat\mu(My)$.
Therefore,
\[
\bar\mu \times (\hat\mu \times \rho)\;(y,z) = \frac{\hat\mu^0(My)
\hat\mu^1(Ez)}{|My||Ez|}
\]
which proves the identity (\ref{ur}).

The representation (\ref{ur}) enables us to consider the initial
distribution as constructed directly from the $\bar\mu$ by
randomizing it with the {\it a priori} distribution $\hat\mu
\times \rho$ which is not time-invariant under the dynamics and
depends on the initial state of the environment, cf. the first
remark at the end of Section \ref{sec: er}. The probability of a
trajectory $\bar\om = (\bar\om_0,\ldots,\bar\om_n)$ in $\bar\Om$,
i.e., of the system, may then be evaluated as follows:
\begin{equation}\label{q}
\begin{split}
  \bsP_{\hat\mu^0\otimes\hat\mu^1}(\bar\om) 
    &\equiv (\bar\mu \times (\hat\mu \times \rho))
    (\cap_{j=0}^{n} f^{-j} \bar\om_j)
  = \frac{\bar\mu(\bar\om_0)}{\hat\mu(\bar\om_0)}
    (\hat\mu \times \rho) (\cap_{j=0}^{n} f^{-j} \bar\om_j)
\\
  &= \frac{\bar\mu(\bar\om_0)}{\hat\mu(\bar\om_0)} \sum_{\bar p(\hat\om) = \bar\om}
  \bsP_{\hat\mu}(\hat\om)
\end{split}
\end{equation}
where $\bsP_{\hat\mu}(\hat\om)$ is the probability of the
trajectory $\hat\om$ on $\hat\Om$  started from $\hat\mu$:
\begin{equation}\nonumber
  \bsP_{\hat\mu}(\hat\om)
  \equiv (\hat\mu \times \rho)(\cap_{j=0}^{n} f^{-j} \hat\om_j)
\end{equation}
and we sum in (\ref{q}) over all trajectories on $\hat\Omega$
(i.e., for environment and system together) that coincide in their
first coordinate with the given trajectory $\bar\om$.  Similarly,
for the time-reversed trajectory $\Th\bar\om$ with the system's
initial distribution $\bar\nu\pi$ we have the probability (the
initial microstates sampled from the distribution $\bar\nu\pi
\times (\hat\mu\pi \times \rho) = (\hat\nu^0\pi \otimes
\hat\mu^1\pi)\times\rho$)
\begin{equation}\label{qi}
\begin{split}
  \bsP_{\hat\nu^0\pi\otimes\hat\mu^1\pi}(\Th\bar\om) &=
    \frac{\bar\nu(\bar\om_n)}{\hat\mu(\bar\om_n)}
    \sum_{\bar p(\hat\om) = \Th\bar\om} \bsP_{\hat\mu\pi}(\hat\om)
\\
 & = \frac{\bar\nu(\bar\om_n)}{\hat\mu(\bar\om_n)} \sum_{\bar p(\hat\om) = \bar\om}
    \frac{\hat\mu(\hat\om_n)}{\hat\mu(\om_0)}\frac{|\hat\om_0|}{|\hat\om_n|}
   \bsP_{\hat\mu}(\hat\om)
\end{split}
\end{equation}
where we used a version of (\ref{prop1}). As always, we want to
take the ratio of (\ref{q}) and (\ref{qi}). We write this out in
the most explicit form:
\begin{equation}\label{expl}
\frac{\bsP_{\hat\mu^0\otimes\hat\mu^1}(\bar\om)}
{\bsP_{\hat\nu^0\pi\otimes\hat\mu^1\pi}(\Th\bar\om)} =
\frac{\hat\mu^0(M_0)}{\hat\nu^0(M_n)} r_n^{-1}(\bar\om)
\end{equation}
where
\[
r_n(\bar\om)\equiv \frac{\sum_{E_0,\ldots,E_n}
\frac{\hat\mu^1(E_n)}{\hat\mu^1(E_0)}
\frac{|M_0||E_0|}{|M_n||E_n|}
\bsP_{\hat\mu}[(M_0,E_0),\ldots,(M_n,E_n)]}{\sum_{E_0,\ldots,E_n}
  \bsP_{\hat\mu}[(M_0,E_0),\ldots,(M_n,E_n)]}
\]
for a trajectory $\bar\om=(M_0,\ldots,M_n)$ of the system and the
sums are over trajectories $(E_0,\ldots,E_n)$ of the environment.

The identity (\ref{expl}) is still exact and general. To proceed,
we choose first to be more specific about the nature of the
environment.  As example, we suppose that the environment consists
of $m$ heat baths which are taken very large and which are
prepared at inverse temperatures $\beta_1,\ldots,\beta_m$. This
means that $\hat\Om^1$ and $\hat\mu^1$ are split further as
$m-$fold products and that the trajectories of the environment are
obtained from the successive values of the energies $E_i^k$ at
times $i\delta$ in all heat baths, $k=1,\ldots,m$.  We also
suppose that these reservoirs are spatially separated, each being
in direct contact only with the system.  Through these system-heat
bath interfaces a heat current will flow changing the energy
contents of each reservoir.  It implies that even though the
initial energies $E_0^k$ are sampled from $\hat\mu^1$ giving
inverse temperatures $\beta_k$ to each of the heat baths, {\it a
priori} it need not be that the final energies $E_n^k$ can be
considered as sampled corresponding to the same temperatures. At
this moment we need a steady state assumption for the reservoirs:
that they maintain the same temperature during the evolution, or
more precisely,

\vspace{2pt}\noindent {\bf A1:} The path space measure
$\bsP_{\hat\mu}$ gives full measure to those trajectories for
which
\[
|E_n^k - E_0^k| \leq  o(\sqrt{V})
\]
of the order of less than the square root of the size of the
volume $V$ of the
environment. \\

For simplicity, we have used here $o(\sqrt{V})$ as a safe
estimate. One could assume much less: the energy differences
$|E_n^k - E_0^k|$ will be of the order of the product of the
surface area $\partial_k \Lambda$ through which heat will flow
between the $k-$th  heat bath and the system, and the heat current
(i.e., the flow of energy per unit area and per unit time) and the
time
$t=n\delta$.  One can recognize this from the calculations in Appendix B.\\
  We need this
assumption A1 to get rid of the ratio
$\hat\mu^1(E_n)/\hat\mu^1(E_0)$ in (\ref{expl}). Under A1, this
ratio is essentially equal to one, when the initial dispersion of
the energy values under $\hat\mu^1$ is much larger than the
changes of energy. For simplicity (but without loss of generality)
we can suppose that for each heat bath of spatial size $V$,
$\hat\mu^1$ gives the uniform distribution over an interval
$[\caE^{k} - \ve, \caE^{k} + \ve]$ where we take $\caE^{k} =
\caO(V)$ and $\ve = o(V)$. This is just the simplest
representation for an initial distribution that is peaked at
energy value $\caE$ with deviations of order $\ve = \sqrt{V}$ say.
Within such an interval, the temperatures $1/\be_{k}$ are
essentially constant if the size of the interval is large compared
with the possible changes of the energy due to the flows from the
system. These statements become sharp in the limit $V \to \infty$
but for finite reservoirs this steady state assumption can be
expected realized over times $t \leq t^\star$ with $t^\star =
t^\star(V)$ growing with $V$ less than as $\sqrt{V}$. As
conclusion, we set $\hat\mu^1(E_n)/\hat\mu^1(E_0) =1$ in
(\ref{expl}).

For notation, we denote by
\begin{equation}\label{cbs}
  \hat{S}_B^0(M_n) - \hat{S}^0_B(M_0) = \ln\frac{|M_n|}{|M_0|}
\end{equation}
the change of Boltzmann entropy of the system.  Usually, at least
in close to equilibrium treatments, this change is divided into two parts:
one corresponding to the entropy production properly speaking and one term 
corresponding to the entropy current through the surface of the system; 
one refers to it as an entropy balance equation, see Appendix D for 
a short discussion. The latter, the entropy current, is responsible for 
the change of entropy in the reservoirs, here
\begin{equation}\label{cbr}
\ln \frac{|E_n|}{|E_0|} = \sum_{k=1}^m [\hat{S}_B^k(E_n^k) -
\hat{S}_B^k(E_0^k)]
\end{equation}
the sum of changes of the Boltzmann entropy in each bath. The sum
of (\ref{cbs}) and (\ref{cbr}) is the total change of entropy
(where total refers to the closed system consisting of the
(smaller) system and all the reservoirs) and thus equals the
entropy production.  We write it as
\begin{eqnarray}\label{dse}
&& \De S_B (\bar\om) \equiv \hat{S}_B(M_n) - \hat{S}_B(M_0)
\\\nonumber && - \ln
\frac{\sum_{E_0,\ldots,E_n} e^{-\sum_{k=1}^m [\hat{S}_B^k(E_n^k) -
\hat{S}_B^k(E_0^k)]}
\bsP_{\hat\mu}[(M_0,E_0),\ldots,(M_n,E_n)]}{\sum_{E_0,\ldots,E_n}
\bsP_{\hat\mu}[(M_0,E_0),\ldots,(M_n,E_n)]}
\end{eqnarray}
As in (\ref{totalen}), it is natural to take  $\hat\nu^0 =
\hat\mu_t^0 \equiv \bar p(\bar\mu \times (\hat\mu \times \rho))_t$
in (\ref{expl}) which corresponds to the projection at time
$t=n\delta$ on the system,  and we obtain a first

\vspace{2pt} \noindent {\bf Analogue of Proposition \ref{cor}:} \\
\begin{equation}\label{cors}
\ln \frac{\bsP_{\hat\mu^0\otimes\hat\mu^1}(\bar\om)}
{\bsP_{\hat\mu^0_t\pi\otimes\hat\mu^1\pi}(\Th\bar\om)}
 =
\De S_B (\bar\om) + [-\ln \hat\mu^0_t(\bar\om_n) + \ln
\hat\mu^0(\bar\om_0)]
\end{equation}   \\

We can still do  better by reconsidering (\ref{dse}), in
particular the expectation over the path-space measure of the
exponential change of Boltzmann entropies in the heat baths. These
changes are caused by the heat dissipated in each of the
reservoirs and it therefore corresponds to the entropy current.
Since this is the energy current divided by the temperature of the
reservoir, it should be possible to express it directly in terms
of the microscopic trajectory over the surface separating the heat
bath from the system.  The basic question is now in what sense the
trajectory $\bar\om$ of the system determines the trajectory
$\hat\om$ of the total system. In the context of Hamiltonian
dynamics it is not hard to see that (again, provided that the
reservoirs are coupled to different parts of the system) the
trajectory $\hat\om$ is uniquely determined by its projection onto
the system, $\bar\om$, and by the initial energies of the
reservoirs, $E^{k}_0$. We formulate this in the form of another
assumption: \\

\vspace{2pt} {\bf A2:} Let $\hat\om$ and $\hat\om'$  be two
trajectories of the total system such that
$\bar{p}(\hat\om)=\bar{p}(\hat\om')=\bar\om$. Then, for typical
trajectories obtained as successive reduced states from the
Hamiltonian dynamics,
the energy changes $E^k_n - E^k_0$ depend only on $\bar\om$. \\

This should be understood in the sense of all allowed
trajectories, typical here referring to the path-space measure
$\bsP_{\hat\mu}$.   This assumption is  too strong. The problem is
that we also consider reduced states on the level of the system
itself and that the $\bar\om$ is not the continuous time
microscopic trajectory of the system. It would for example have
been better to use a finer time-scale for the evolution of the
reduced states of the system (compared with that of the
reservoirs). One could remedy that but we prefer to
stick to A2 for simplicity, see Appendix B.\\
Assuming A2, we already know that $\De E^{k} \equiv E^{k}_n - E^{k}_0$ 
only depends on $\bar\om: \De E^{k}= \De E^{k}(\bar\om)$. So, the
change of Boltzmann entropy in each reservoir,
$\hat{S}_B^{k}(E^{k}_n) - \hat{S}_B^{k}(E^{k}_0) =
\hat{S}_B^{k}(E^{k}_0 + \De E^{k}(\bar\om)) -
\hat{S}_B^{k}(E^{k}_0)$, depends on $\bar\om$ and $E^{k}_0$. In
order to get rid of the dependence on the initial state of the
reservoirs, we must again use the distribution  $\hat\mu^1$ and
that the reservoirs are large. After all, thermodynamic behavior
implies that the $E_i^{k} = \caO(V)$ and $\hat{S}_B^{k}(E_i^k) =
\caO(V)$, all of the order of the volume $V$ of the reservoirs,
while $\be_{k} \equiv
\partial \hat{S}_B^{k}(E_i^k) /
\partial E_i^{k}$ is kept fixed.
This is again our steady state assumption A1 for the environment;
the reservoirs are heat baths at a fixed temperature.  The change
of Boltzmann entropy in each of the reservoirs is then
\begin{equation}\nonumber
  \hat{S}_B^{k}(E^{k}_n) - \hat{S}_B^{k}(E^{k}_0)
  = \be_{k} \De E^{k}(\bar\om) + \caO(\frac 1{\sqrt{V}})
\end{equation}
for all trajectories of the system $\bar\om$ and essentially all
initial energies $E^{k}_0 \in (\caE^{k} - \ve, \caE^{k} + \ve)$
(most of trajectories started inside the interval will not leave
it).     This implies that the total change of entropy appearing
in (\ref{dse}) and in (\ref{cors}) is, in good approximation,
\begin{equation}\label{de}
\De S_B(\bar\om) = \hat{S}_B^0(\bar\omega_n) -
\hat{S}_B^0(\bar\om_0) + \sum_k \be_{k} \De E^{k}(\bar\om)
\end{equation}
What we have gained with respect to (\ref{dse})-(\ref{cors}) is
that this variable entropy production only depends on $\hat\mu^1$
through the initial temperatures of the reservoirs.\\
We denote
\begin{equation}\label{nota}
R_{\hat\mu^0}^{\hat\mu^1,t}(\bar\om) \equiv \ln
\frac{d\bsP_{\hat\mu^0\otimes\hat\mu^1}}
{d\bsP_{\hat\mu^0_t\pi\otimes\hat\mu^1\pi} \Th}(\bar\om)
\end{equation}
We conclude with the final result obtained under the assumptions
above:

\vspace{2pt}\noindent {\bf Analogue of Proposition \ref{cor}:}
\begin{equation}\label{sug}
R_{\hat\mu^0}^{\hat\mu^1,t}(\bar\om)
= \hat{S}_B^0(\bar\omega_n) -
\hat{S}_B^0(\bar\om_0) + \sum_k \be_{k} \De E^{k}(\bar\om)+ [-\ln
\hat\mu_t^0(\bar\om_n) + \ln \hat\mu^0(\bar\om_0)]
\end{equation}                                               \\

There are two big modifications with respect to (\ref{totalen})
for closed systems.  First, it is important to remember that the
$\De E^{k}(\bar\om)$ in (\ref{sug}) are in general not differences
of the form $E_n^k(\bar\om_n) - E_0^k(\bar\omega_0)$ but they
represent the heat flow depending on the complete path $\bar\om$.
So the right-hand side of (\ref{sug}) is not a difference.
Secondly, here it is very well possible to take
$\hat\mu_t^0=\hat\mu^0$ at least for small enough times $t$
(compared to $\sqrt{V}$) what refers to the full steady state. In
other words, we can study the stationary regime where the
distribution $\hat\mu^0$ of the system is time-invariant.

\vspace{2pt}\noindent {\bf Remark 1:} The same construction
applies of course also  when the system is coupled to only one
reservoir or to various reservoirs at the same temperature
$\beta^{-1}$. We take $\hat\Om^0 = \Om^0=\bar\Om$ so that the
system is described via its microscopic states $y$. The trajectory
$\bar\om$ gives successive microscopic states $\bar\om =
(y_0,\ldots,y_n)$ and the first two terms on the right-hand side
of (\ref{sug}) are identically zero. By energy conservation, the
total change of energy $\sum_k (E_n^k - E_0^k) = \De E$ in the
reservoirs is always of the form $\De E = H(y_0) - H(y_n)$, the
difference of the initial and final energies of the system.
Therefore, in (\ref{de}),
\[
\De S_B(\bar\omega) = \beta [H(y_0) - H(y_n)]
\]
It is interesting to see that then, when taking $\hat\mu^0(y) =
\hat\mu_t^0(y) \sim \exp[-\beta H(y)]$ a Gibbs measure at inverse
temperature $\beta$, the expression (\ref{sug}) becomes zero.

\vspace{2pt}\noindent {\bf Remark 2:}  The same construction
applies also to other scenario's (instead of via heat reservoirs)
but it needs some change in notation.  As an example of another
physical mechanism we can consider the system coupled to a heat
bath at constant temperature $\beta^{-1}$ where some parameters
(e.g. interaction coefficients) in the interaction of the
components of the system are changed. This means that the
effective Hamiltonian $H(\tau) \equiv H(\lambda(\tau),y), \tau\in
[0,t],$ of the system is time-dependent with final value $H_f(y)
\equiv H(\lambda(t),y)$ and initial value $H_i(y) \equiv
H(\lambda(0),y)$. To change the parameter $\lambda$ from
$\lambda(0)$ to $\lambda(t)$ some heat must flow from the bath
into the system so that the change of entropy of the bath equals
$\De S_B = -\beta[H_f( y_n) - H_i(y_0) - W_t]$ where
 $W_t$ is the work
done over the time $[0,t]$. If we assume that the initial
distribution $\hat\mu^0 = \exp[-\beta H_i]/Z_i$ and the final
distribution $\hat\mu_t^0 = \exp[-\beta H_f]/Z_f$ are describing
equilibrium with respect to the Hamiltonians $H_i$ and $H_f$
respectively, then (\ref{sug}) becomes
\begin{equation}\label{sugwo}
R_{\hat\mu^0}^{\hat\mu^1,t}(\bar\om)
= \beta W_t(\bar\omega) -\beta \De F
\end{equation}
where $\De F \equiv -\beta^{-1}[\ln Z_f - \ln Z_i]$ is the change
of (equilibrium) Helmholtz free energy.\\
This and the previous remark also indicate that the physical
significance of the terms 
$-\ln \hat\mu_t^0(\bar\om_n) + \ln \hat\mu^0(\bar\om_0)$ 
in (\ref{sug}) depends on what can
physically be assumed or said about $\hat\mu^0$ and $\hat\mu_t^0$.
This can be different from case to case. Yet here again, as
already said in Remark 2 of the previous section, while these
terms have {\it a priori} nothing to do with entropy production,
adding them gives rise to a more convenient form, both for the
properties of the average (mean entropy production) and for the
fluctuations of the entropy production.

\vspace{2pt} The mean entropy production rate in the steady state
where $\hat\mu_t^0=\hat\mu^0\equiv \bar\mu$ is time-invariant is
obtained from taking the average of (\ref{sug}) with respect to
$\bsP_{\hat\mu^0\otimes\hat\mu^1}$. Let $\bsE_{\bar\mu}$ stand for
the expectation (we do not indicate the dependence on
$\hat\mu^1$).  As is the case for our example, we suppose for
simplicity for the rest of this section that $\hat\mu^1 =
\hat\mu^1\pi$.  We have then the

\vspace{2pt} \noindent {\bf Steady state analogue of Proposition
\ref{propo2}:} The entropy production $\De S_B$ of (\ref{de})
satisfies
\begin{equation}\label{exprz}
\bsE_{\bar\mu}[e^{- z \De S_B}]= \bsE_{\bar\mu} [e^{-(1- z) \De
S_B} \frac{\bar\mu(\pi\om_n)}{\bar\mu(\om_0)}]
\end{equation}
for all complex numbers $z$.  In particular, its expectation
equals the mean entropy current $=$ mean entropy production $=$
\begin{equation}\label{gepc}
\bsE_{\bar\mu}[\De S_B]= \sum_k \be_{k} \bsE_{\bar\mu}[ \De E^{k}]
\geq 0
\end{equation}

The relation (\ref{exprz}) expresses a symmetry in the
fluctuations of $\De S_B$.  Modulo some technicalities that amount
to estimating space-time boundary terms, as explained in
\cite{M,mm}, it reproduces almost immediately the Gallavotti-Cohen
symmetry, \cite{GC}. While it is the theory of smooth dynamical
systems that has guided us to it, in our analysis, nothing has
remained of a chaoticity hypothesis.\\ The relation (\ref{gepc})
states the positivity of the mean entropy production. From its
proof (below) we can understand under what (nonequilibrium)
conditions, it is in fact strictly positive.

 The basic identity
that drives fluctuation-symmetry relations is
\begin{equation}\label{bid}
\bsE_{\bar\mu}[ e^{-z \cal{R}(\bar\om)} \psi(\om)] =
\bsE_{\bar\mu}[e^{-(1-z) \cal{R}(\bar\om)} \psi(\Th\om)]
\end{equation}
for every function $\psi$ of the trajectory of the system and with
\[
\cal{R}(\om) \equiv \ln \frac{d\bsP_{\bar\mu\otimes\hat\mu^1}}
{d\bsP_{\bar\mu\otimes\hat\mu^1} \Th}(\bar\om)
\]
This identity (\ref{bid}) follows from the very definition of
$\cal{R}$ as the logarithmic ratio of two probabilities from which
also $\cal{R}(\Th\bar\om) = - \cal{R}(\bar\om)$. The equation
(\ref{exprz}) follows simply by taking for $\psi$ in (\ref{bid}),
$\psi(\bar\om) =
[\bar\mu(\bar\omega_0)/\bar\mu(\pi\bar\om_n)]^z.$\\ Before we give
the proof of (\ref{gepc}), we give the version for the transient
regime of the system (steady state for the reservoirs):

\vspace{2pt} \noindent {\bf Transient regime analogue of
Proposition \ref{propo2}:} Recall the notation (\ref{nota}). Then,
\begin{equation}\label{exprzt}
\bsE_{\hat\mu^0\otimes\hat\mu^1}[e^{- R_{\hat\mu^0}^{\hat\mu^1,t}}] = 1
\end{equation}
Its expectation equals
\begin{equation}\label{gepct}
\bsE_{\hat\mu^0\otimes\hat\mu^1}[R_{\hat\mu^0}^{\hat\mu^1,t}]=
S_G(\hat\mu_t^0) - S_G(\hat\mu^0) + \sum_k \be_{k}
\bsE_{\hat\mu^0\otimes\hat\mu^1}[ \De E^{k}] \geq 0
\end{equation}

The relation (\ref{exprzt}) for the example (\ref{sugwo}) gives
the irreversible work - free energy relation of Jarzynski,
\cite{jar}.\\
Now to the proofs of (\ref{gepc})-(\ref{gepct}).  As before in
(\ref{expr}), (\ref{exprzt}) just expresses a normalization of the
probability measure $\bsP_{\hat\mu_t^0 \pi\otimes\hat\mu^1}\Th$.
The equality in (\ref{gepct}) follows as in (\ref{gep}) from
taking the expectation of (\ref{sug}). From the Jensen inequality
applied to (\ref{exprzt}), we obtain the inequality in
(\ref{gepct}).  The relation (\ref{gepc}) now follows from
applying stationarity $\hat\mu^0=\hat\mu_t=\bar\mu$.\\
We thus see that the positivity in (\ref{gepc}) and in
(\ref{gepct}) follows from convexity.  By the same argument, the
strict positivity will express that the two path space measures
$\bsP_{\hat\mu^0\otimes\hat\mu^1}$ and
$\bsP_{\hat\mu^0\otimes\hat\mu^1}\Th$ are really different, i.e.,
applying time-reversal really has an effect. (In \cite{M} this is
expressed via the relative entropy between these two path space
measures.)

\noindent {\bf Remark 3:} Note that the above fluctuation
identities (\ref{exprz}), (\ref{bid}) and (\ref{exprzt}) do not
depend on Assumption A2.  We can repeat them directly starting
from (\ref{cors}).

\section{Markov approximation}\label{sec:markov}

The stochastic processes of the previous sections give the
statistics of trajectories for reduced states induced by the
Hamiltonian dynamics.  The stochasticity does not represent
microscopic or intrinsic randomness, whatever that means, and is
not an easy substitute for chaoticity.  In the present section we
make an approximation for this stochastic evolution that does go
in the direction of assuming some chaoticity but again on the
level of reduced states.

\subsection{Closed systems}
We refer here to Section \ref{sec: er}. Look at the time-evolved
measure $(\hat\mu\times\rho)_t$ starting from $\hat\mu\times\rho$
at time zero: for $t=n\delta$,
\[
(\hat\mu\times\rho)_t(x) = \hat\mu\times\rho (f^{-n}x)
\]
Remember that we have used before its projection $\hat\mu_t$ on
$\hat\Gamma$.
 Observe now that, quite generally,
$(\hat\mu\times\rho)_t \neq \hat\mu_t\times\rho$. That is: the
phase space distribution does not remain microcanonical; when two
points $x,y\in \Gamma$ fall into the same reduced state ($Mx=My$),
it need not be that
$(\hat\mu\times\rho)_t(x)=(\hat\mu\times\rho)_t(y)$. This is an
instance of so called memory-effects; the process $\bsP_{\hat\mu}$
does certainly not correspond to
a Markov process on $\hat\Gamma$.\\
We can obtain a Markovian approximation by forcing uniformization
at each step in the evolution. We then define the discrete time
Markov approximation via the updating
\begin{equation}\label{markov}
\tilde\mu_n = p((\tilde\mu_{n-1}\times\rho)_\delta),\,n=1,2,\ldots
\end{equation}
or more explicitly, from (\ref{ratio}),
\[
\tilde\mu_n(M) = \sum_{M'\in\hat\Gamma} \tilde\mu_{n-1}(M')
\frac{|f^{-1}M\cap M'|}{|M'|}
\]
corresponding to the Markov chain on $\hat\Gamma$ with transition
probabilities $p(M',M) = |f^{-1}M\cap M'|/|M'|$.  Naturally it
satisfies the detailed balance condition
\begin{equation}\label{db}
|M|\, p(M,M') = |M'|\, p(\pi M',\pi M)
\end{equation}
or
\begin{equation}\label{db1}
\frac{p(M,M')}{p(\pi M',\pi M)} = e^{\hat{S}_B(M') - \hat{S}_B(M)}
\end{equation}
It is an approximation in the sense that the evolution defined by
(\ref{markov}) corresponds to a repeated randomization of the
`true' evolution.  We expect it to be a good approximation in so
far that $|M\cap f^j M'| \simeq |M||M'|/|\Gamma|$.  That is to
say, for $\delta$ large enough for the averaging over the reduced
state to be valid. That is a mixing condition but for the
evolution over the reduced states (as for Gibbs' inkdrop), see
\cite{bric} for similar remarks. It also implies relaxation to
equilibrium. Usually however this is combined with other limiting
procedures  through which the reduced variables (or their
fluctuations) get an autonomous (stochastic) evolution. Most
important in all this however remains the `proper choice' of
reduced states (or, thermodynamic variables).

We now have a Markov chain $(X_k)$ on $\hat\Gamma$ with transition
probabilities $p(M,M')$ and
\[
\prob_{\tilde\mu}[X_n=M_n,\ldots,X_0=M_0]=
\tilde\mu(M_0)p(M_0,M_1)\ldots p(M_{n-1},M_n)
\]
is the probability of a trajectory $\omega=(M_0,M_1,\ldots,M_n)\in
\hat\Gamma^{n+1}$ when the Markov chain was started from the
probability measure $\tilde\mu$. We have instead of (\ref{prop}):
\begin{multline}\label{propmarkovdb1}
  \frac{\prob_{\tilde\mu}(X_n = M_n,\ldots,X_0 = M_0)}
  {\prob_{\tilde\nu\pi}(X_n =\pi M_0,\ldots,X_0 = \pi M_n)} =
\\
  \frac{\tilde\mu(M_0)}{\tilde\nu(M_n)}\exp[\sum_{k=0}^{n-1}\ln
  \frac{p(M_k,M_{k+1})}{p(\pi M_{k+1},\pi M_k)}]
\end{multline}
Upon substituting (\ref{db}), the exponential in
(\ref{propmarkovdb1}) equals $|M_n|/|M_0|$ and, perhaps
surprisingly, the identity (\ref{prop})-(\ref{prop1}) is
unaffected in the Markov approximation:
\begin{equation}\label{propmarkovdb}
\frac{\prob_{\tilde\mu}(X_n = M_n,\ldots,X_0 = M_0)}
{\prob_{\tilde\nu\pi}(X_n =\pi M_0,\ldots,X_0 = \pi M_n)} =
\frac{\tilde\mu(M_0)}{\tilde\nu(M_n)}\frac{|M_n|}{|M_0|}
\end{equation}
Furthermore, take now $\tilde\nu= \tilde\mu_n$ of (\ref{markov})
and let us denote as in Proposition \ref{propo2},
\[
R_{\tilde \mu}^n(\omega)\equiv \ln \frac{\prob_{\tilde\mu}(X_n =
M_n,\ldots,X_0 = M_0)} {\prob_{\tilde\mu_n\pi}(X_n =\pi
M_0,\ldots,X_0 = \pi M_n)}
\]
Its expectation, as in (\ref{gep}), under the now Markovian path
space measure $\bsP_{\tilde\mu}$ is
\begin{equation}\label{sep}
\bsE_{\tilde\mu}[R^n_{\tilde\mu}]=\sum_{\omega\in
\hat\Gamma^{n+1}} \bsP_{\tilde\mu}(\omega) R_{\tilde
\mu}^n(\omega) =  S(\tilde\mu|\tilde\rho) -
S(\tilde\mu_n|\tilde\rho) \geq 0
\end{equation}
the difference of relative entropies with respect to the
stationary (reversible) probability measure $\tilde\rho(M)\equiv
|M|/|\Gamma|$; the relative entropy is defined from
$S(\tilde\nu|\tilde\rho)\equiv \sum_M \tilde\nu(M) \ln
\tilde\nu(M)/\tilde\rho(M)$. The identities (\ref{sep}) and
(\ref{gep}) are consistent since the Gibbs entropy can be written
in terms of this relative entropy as $S_G(\tilde\nu)= \ln |\Gamma|
- S(\tilde\nu|\tilde\rho)$.

\subsection{Open systems}
We refer here to Section \ref{sec: cou}. In the same spirit as
above, we get the Markov approximation for open systems by
 following the procedure of Section \ref{sec: cou}.  We now get Markov
processes with transition probabilities
\[
q(M,M') = \bsP_{M\otimes\hat\mu^1}(M,M')
\]
where we have understood $\hat\mu^0 = \delta(M -\cdot)$. We will
again suppose for the enviroment that $\hat\mu^1\pi = \hat\mu^1$.
These transition probabilities then satisfy, from (\ref{sug}),
\begin{equation}\label{ldba}
\frac{q(M,M')}{q(\pi M', \pi M)} = \exp \De S_B(M,M')
\end{equation}
The measures $\tilde\mu$ of above now correspond to the
distribution of the internal degrees of freedom (the open system).
The important change is  that detailed balance may be violated
from the action of reservoirs maintained at different but fixed
temperatures or chemical potentials. We can for example substitute
(\ref{de}) in (\ref{ldba}) to retain only a local detailed balance
condition, that is
\[
\frac{q(M,M')}{q(\pi M', \pi M)} =
 \exp[\hat{S}_B^0(M') -
\hat{S}_B^0(M) + \sum_k \be_{k} \De E^{k}(M,M')]
\]
Depending on the transition $M\rightarrow M'$, in particular,
where this transition  of the state of the system is localized,
various terms in the exponential can become zero or non-zero,
 see also
(\ref{ldb}) below.\\  While the formal structure of the Markov
approximation for open systems runs exactly similar to what we did
for closed systems, cf. (\ref{markov}), we remark that its
validity now requires more than what was mentioned following
(\ref{db1}).  In fact, a competing requirement enters if we wish
to maintain assumption A2 of the previous section.  Assumption A2
will be more reliable in so far as the $\delta$ (i.e., the time
steps in the trajectory of reduced states) is smaller while the
mixing condition on the level of reduced states that justifies the
Markov approximation requires large enough $\delta$.  Again, as
mentioned following assumption A2, this motivates using different
time scales for the evolution of the reduced states in system and
enviroment.

We now have a Markov chain $(X_k)$ on $\hat\Om^0$ with transition
probabilities $q(M,M')$, and
\[
\prob_{\tilde\mu}[X_n=M_n,\ldots,X_0=M_0]=
\tilde\mu(M_0)q(M_0,M_1)\ldots q(M_{n-1},M_n)
\]
is the probability of a trajectory
$\bar\omega=(M_0,M_1,\ldots,M_n)$ when the Markov chain was
started from the probability measure $\tilde\mu$, we have instead
of (\ref{prop}):
\begin{multline}\label{propmarkov}
  \frac{\prob_{\tilde\mu}(X_n = M_n,\ldots,X_0 = M_0)}
  {\prob_{\tilde\nu\pi}(X_n =\pi M_0,\ldots,X_0 = \pi M_n)} =
\\
  \frac{\tilde\mu(M_0)}{\tilde\nu(M_n)}\exp[\sum_{k=0}^{n-1}\ln
  \frac{q(M_k,M_{k+1})}{q(\pi M_{k+1},\pi M_k)}]
\end{multline}
As motivated in Section \ref{sec: cou}, its logarithm will
continue to interest us as variable entropy
production.\\

From (\ref{propmarkov}), we see that the variable entropy
production is now given by:
\begin{equation}\label{br}
 \sum_k \ln \frac{q(M_k,M_{k+1})}{q(\pi M_{k+1},\pi
M_k)}
\end{equation}
Furthermore, for open systems, the relation (\ref{sep}) gets
replaced with
\begin{eqnarray}\label{sep1}
\lefteqn{\bsE_{\tilde\mu}[R^n_{\tilde\mu}] = }
\\\nonumber
  & & S(\tilde\mu_n) - S(\tilde\mu) + \sum_{k=0}^{n-1} \sum_{M,M'}
  \tilde\mu_k(M) q(M,M') \ln \frac{q(M,M')}{q(\pi M',\pi M)} \geq 0
\end{eqnarray}
and there is in general no way to write this as a change in
relative entropies $S(\tilde\mu|\tilde\rho) -
S(\tilde\mu_n|\tilde\rho)$. In other words, in general, there is
no role for the time-derivative of the relative Shannon entropy as
total entropy production. When $\tilde\mu=\tilde\mu_n$ is
stationary for the Markov chain, the right-hand side of the
equality in (\ref{sep1}) gives us the mean entropy production rate
as
\begin{equation}\label{mep}
\sum_{M,M' \in \hat\Gamma} \tilde\mu(M) q(M,M') \ln
\frac{q(M,M')}{q(\pi M',\pi M)}
\end{equation}
which (up to the inclusion of the time-reversal involution $\pi$)
is the standard expression for an effective Markovian dynamics
modeling a nonequilibrium steady state, see e.g. \cite{Schn,ELS}.
Note  that if $\tilde\mu$ is stationary under updating with
transition probabilities $q(M,M')$, then $\tilde\mu\pi$ is
stationary under updating with the transition probabilities
$\Theta q(M,M')\equiv q(\pi M', \pi M) \tilde\mu(\pi
M')/\tilde\mu(\pi M)$ for the time-reversed process.  It is then
easy to see that the mean entropy production rate is positive and
equal for both stationary processes. Or, the mean entropy
production is time-reversal invariant.  This again is ultimately a
consequence of the dynamic reversibility of the microscopic
dynamics and it yields interesting
by-products (like Onsager reciprocities) as discussed in \cite{Liu}.\\
For the pathwise expression of the entropy production rate, we
look back at (\ref{br}).  The entropy production per time-step is
\begin{equation}\label{brr}
\sigma_B^n(\omega) \equiv \frac 1{n} \sum_{k=0}^{n-1} \ln
\frac{q(\omega_k,\omega_{k+1})}{q(\pi \omega_{k+1},\pi \omega_k)}
\end{equation}
Note again that $\sigma_B^n(\Theta\omega) = -\sigma_B^n(\omega)$
and that, when $\tilde\mu$ is stationary,
\[
R_{\tilde\mu}^n(\omega)/n = \frac 1{n}
\ln\frac{\tilde\mu(\omega_0)}{\tilde\mu(\omega_n)} +
\sigma_B^n(\omega)
\]
This leads again as in (\ref{exprz}) and in (\ref{bid}) almost
directly to a Gallavotti-Cohen symmetry, \cite{GC,M}.

For a continuous time Markov chain $(X_t)$ on a finite set with
transition rates $k(M,M')$, similarly, the entropy production rate
in the distribution $\mu$ is given by
\begin{equation}\label{contrate}
\sigma(\mu)\equiv \frac 1{2}\sum_{M,M'} [k(M,M')\mu(M) -
k(M',M)\mu(M')]\ln\frac{k(M,M')\mu(M)}{k(M',M)\mu(M')}
\end{equation}
(We have set $\pi=$ identity for simplicity.)  Let us take
\[
 k(M,M') =  k_0(M,M') e^{\varepsilon \psi(M,M')/2}
\]
 where $k_0$ is the rate for a detailed balance evolution with unique reversible measure
 $\mu_0$. We
assume that there is a unique stationary measure $\mu_\varepsilon$
with $\varepsilon$ measuring the distance from equilibrium:
\begin{equation}\label{ldb}
\frac{k(M,M')}{k(M',M)} = \frac{\mu_0(M')}{\mu_0(M)}
\exp[\varepsilon \psi^{\text{as}}(M,M')]
\end{equation}
Here $\psi^{\text{as}}(M,M') = -\psi^{\text{as}}(M',M)=
[\psi(M,M') - \psi(M',M)]/2$ originates in some driving. We did
not indicate it but the $k(M,M')$ and therefore the functional
$\sigma$ in (\ref{contrate}) depend now on $\varepsilon$.  One can
then check that $\sigma(\mu)$ is minimal for a probability measure
$\mu^\star$ which coincides with $\mu_\varepsilon$ to first order
in $\varepsilon$ (minimum entropy production principle).  A
special case of this calculation can be found in \cite{ELS}.  We
give the general statement and argument in Appendix C.

We next apply the above scheme for a Markov approximation for
closed systems to a diffusion process that appeared in the
Onsager-Machlup paper, \cite{om}.

\section{Application: Gaussian fluctuations}\label{sec: om}

As we have argued before, the entropy production appears as the
source term of time-reversal breaking in the logarithm of the
probability for a preassigned succession of thermodynamic states.
Such calculations were already done to study the fluctuations in
irreversible processes in the work of  Onsager and Machlup in
1953, \cite{om}. Our previous section is some extension of this,
as we will now indicate.

We only redo the very simplest case of \cite{om}, their Section 4
for a single thermodynamic variable $\alpha$ obeying the equation
(in their notation)
\begin{equation}\label{om}
R\dot{\alpha} + s\alpha = \epsilon
\end{equation}
We do not explain here the origin of this equation except for
mentioning that $R$ relates the thermodynamic force to the flux
$\dot{\alpha}$ (assumption of linearity). The constant $s$ finds
its origin in an expansion of the thermodynamic entropy function
$S_T(\alpha) = S_T(0) - s\alpha^2/2$ around equilibrium.  For
every $\alpha, S_T(\alpha)$ is the equilibrium entropy when the
system is constrained to this macroscopic value and can be
identified with the Boltzmann entropy $\hat{S}_B(\alpha)$ (up to
the thermodynamic limit) which is also defined outside
equilibrium. From the expansion of the entropy, the thermodynamic
force $dS_T/d\alpha$ depends linearly on the variable $\alpha$
(Gaussian fluctuations). The right-hand side of (\ref{om}) is
purely random (white noise) with variance $2R$. In this way the
oscillator process 
$d\alpha =  -s/R\; \alpha dt + \sqrt{2/R}\, dW_t$ 
with $W_t$ a standard Wiener process, is obtained for
the variable $\alpha$.\\
The work in \cite{om} is then to calculate the probability of `any
path.' These are the trajectories we had before.  With the current
methods of stochastic calculus, this is not so
difficult.\\
We proceed with (\ref{om}).  Using the path-integral formalism we
can write the `probability' of any path $\omega = (\alpha(\tau),
\tau \in [0,t]) $ with respect to the flat path space `measure'
$d\omega = [d\alpha(\tau)]$:
\[
 \mbox{Prob}_{\tilde\mu}(\omega) \simeq
 \tilde\mu(\alpha(0))\,e^{-\cal{A}(\omega)}
\]
for some initial distribution $\tilde\mu$ and with action
functional
\begin{equation}\label{action}
\cal{A}(\omega) \equiv \frac 1{4} \int_{0}^{t}
R(\dot{\alpha}(\tau) + \gamma \alpha(\tau))^2 d\tau
\end{equation}
 for
$\gamma\equiv s/R$.  There is no problem to make mathematical
sense of this; for example the cross-product
\[
\int_0^t \alpha \dot{\alpha} d\tau  = \int_0^t \alpha \circ
d\alpha
\]
is really a Stratonovich integral and the exponent of the square
$\dot{\alpha}^2$ can be combined with the flat path space measure
to define the Brownian reference measure.  More to the point here
is that the integrand in the action functional $\cal{A}$ can be
rewritten as
\[
R\dot{\alpha}^2(\tau) + \frac {s^2}{R} \alpha^2(\tau) +
\frac{d}{d\tau}(s\alpha^2(\tau))
\]
The last term is minus twice the variable entropy production rate
$\dot{S_T}(\alpha)$. It is the only term in the integrand that is
odd under time-reversal.  So if we take the ratio as in
(\ref{totalen}) but here with $\pi=$ identity, we get, rigorously,
\begin{equation}\label{epom}
\ln \frac{d\bsP_{\tilde\mu}}{d\bsP_{\tilde\mu_t}\Theta}(\omega)=
[S_T(\alpha(t)) - S_T(\alpha(0))] +  [-\ln \tilde\mu_t(\alpha(t))
+ \ln \tilde\mu(\alpha(0))]
\end{equation}
so that, just as in (\ref{totalen}), indeed the change in
thermodynamic entropy is obtained from the source term in the
action functional that breaks the
time-reversal invariance.\\
Onsager and Machlup use the expression (\ref{action}) for the
action functional to derive a variational principle that extends
the so called Rayleigh Principle of Least Dissipation.  The idea
is to take $t$ very small and to seek the $\alpha(t)$ which will
maximize the probability Prob$[\alpha(t)|\alpha(0)]$.  Or, what is
the most probable value of the flux $\dot\alpha$ when you start
from $\alpha(0)$?  This then determines the most probable path.
This means that we should maximize
\[
- \cal{A} = [-\frac 1{4} R \dot{\alpha}^2 - \frac {s^2}{4R}
\alpha^2 + \frac{\dot{S_T}(\alpha)}{2}]t
\]
over all possible $\dot\alpha$, or that we should take
\[
\dot{S_T} - \Phi(\dot\alpha) = \mbox{max.}
\]
where $\Phi(\dot\alpha) \equiv  R \dot{\alpha}^2 /2$ is the so
called dissipation function.  In other words, the maximum (over
the flux) of the difference of the entropy production rate and the
dissipation function determines the most probable path given an
initial value for the thermodynamic variable.  We mention this
here not only because it is a central topic in the Onsager-Machlup
paper but because this Rayleigh principle is often confused with
the minimum entropy production principle that we had at the end of
Section \ref{sec:markov}.  In fact, the Rayleigh principle is more
like a maximum entropy production principle (similar to the Gibbs
variational principle in equilibrium statistical mechanics)
enabling the search for the typical histories.  Of course, its
solution is just (\ref{om}) for $\epsilon=0$, i.e., the
deterministic evolution for the thermodynamic variable, cf.
\cite{ja1}. The minimum entropy production principle on the other
hand, attempts to characterize the stationary states as those
where the entropy production rate is minimal. Both principles have
serious limitations.

\section{Phase space contraction}\label{sec: PSC}

A more recent attempt to model nonequilibrium phenomena that was
largely motivated by concerns of simulation and numerical work,
involves so called thermostated dynamics, see \cite{Dorf,evans}.
These are again as in the previous section, effective models but
now using a deterministic dynamics. First, non-Hamiltonian
external forces are added to the original Hamiltonian equations of
motion to keep the system outside equilibrium.  Since then, energy
is no longer conserved and the system would escape the compact
surface of constant energy, one adds `thermostat forces',
maintaining the energy fixed. There are other possible choices but
they do not matter here. The resulting dynamics no longer
preserves the phase space volume.  We will keep the same notation
as in Section \ref{sec: setup} to denote the discretized dynamics;
$f$ is still an invertible transformation on $\Gamma$ satisfying
dynamic reversibility $\pi f\pi= f^{-1}$ but now the Liouville
measure is not left invariant.  It is important to remember that
$\Gamma$ does no longer represent the phase space of the total
system (subsystem plus reservoirs); it is the phase space of the
subsystem while the action of the environment is effectively
incorporated in $f$.    This environment has two functions at
once: it drives the subsystem in a nonequilibrium state and it
consists of a reservoir in which all dissipated heat
can leak.\\
In the same context it has been repeatedly argued that the phase
space contraction plays the role of entropy production, see e.g.
\cite{ru,GC}.  For thermostated dynamics, there are indeed good
reasons to identify the two and various examples, mostly applied
in numerical work, have illustrated this.  Yet, from a more
fundamental point of view, this needs an argument.  To start,
there is the simple observation that entropy can change in closed
Hamiltonian systems while there is no phase space contraction.
Moreover, even when used for open systems in the steady state
regime, entropy production as commonly understood in irreversible
thermodynamics is  more than a purely dynamical concept. It is
also a statistical object connecting the microscopic complexity
with macroscopic behavior. That was also the reason to introduce
the reduced states and the partitions $\hat\Gamma, \hat\Om$. It is
therefore interesting to see how and when phase space contraction
relates to the concept of entropy production that we have
introduced before.

Since the set-up is here somewhat different from that of Section
\ref{sec: setup}, we denote here the state space by $\cal{M}$
instead of by $\Gamma$. It need not be the set of microstates (as
in thermostated dynamics); it may be the set of possible values
for some hydrodynamic variables, more like our set $\hat\Gamma$.
We think of $\cal{M}$ as a bounded closed and smooth region of
$\bbR^d$. Still, the dynamics $f$ is
assumed dynamically reversible (which would fail for irreversible hydrodynamics).\\
Suppose we have probability densities $\mu$ and $\nu$ on
$\cal{M}$. We replay (\ref{prop}) or (\ref{expl})
but now on the
space $\cal{M}$. For every function $\Phi_n$ on $\cal{M}^{n+1}$,
let $\Phi^*_n$ be given as $\Phi^*_n(x_0,x_1,\ldots,x_n) \equiv
\Phi_n(\pi x_n,\pi x_{n-1},\ldots,\pi x_0)$. We find that
\begin{eqnarray}\nonumber
  \int \Phi^*_n(y,fy,\ldots,f^ny) \nu\pi(y) dy &=&
  \int \Phi_n(\pi f^{n}y,\ldots,\pi y) \nu\pi(y) dy
\\\nonumber
  &=& \int \Phi_n(f^{-n} y,\ldots,y)\nu(y) dy\nonumber
\end{eqnarray}
using dynamic reversibility. Now change variables $y=f^n x$,
\begin{multline}\nonumber
  \int \Phi^*_n(x,fx,\ldots,f^n x) \nu\pi(x) dx =
\\
  \int \Phi_n(x,\ldots,f^n x) \frac{\nu(f^nx)}{\mu(x)}
  |\frac{df^n}{dx}(x)| \mu(x) dx
\end{multline}
or
\[
\langle \Phi^*_n \rangle_{\nu\pi} = \langle \Phi_n r_n
\rangle_\mu, \; r_n(x)=\frac{\nu(f^n
x)}{\mu(x)}|\frac{df^n}{dx}(x)|
\]
This should again be compared with (\ref{prop}) and with
(\ref{expl}). In particular, we see that the phase space
contraction, or more precisely, minus the logarithm of the
Jacobian determinant $- \ln |df^n/dx|$, replaces the total entropy
production (of the total system) we had before:
\begin{equation}\label{repl}
S_B(f^nx)- S_B(x) \longrightarrow - \ln |df^n/dx|
\end{equation}
This requires the dynamical reversibility; without it, even this
purely formal identification is not justified.\\
Looking further to compare with Proposition \ref{cor} and
(\ref{cors}), we can take $\nu(x) = \mu_n(x)$, the time-evolved
density.  Then,
\[
\nu(f^nx) = \mu(x) |\frac{df^{-n}}{dx}(f^nx)|,\quad r_n(x)=1
\]
so that the formal analogue of the right-hand side of
(\ref{totalen}) and (\ref{cors}) now becomes
\begin{equation}\label{pep}
-\ln \mu_n(f^nx) + \ln \mu(x) - \ln |\frac{df^n}{dx}(x)| = 0
\end{equation}
But if we believe in our algorithm for computing the mean entropy
production as in (\ref{gep}) for closed systems and as in
(\ref{gepct}) for open systems,  the expectation of (\ref{pep})
with respect to $\mu$ should give us the mean entropy production;
it remains of course zero:
\begin{equation}\label{peps}
-\int dx \mu_n(x) \ln \mu_n(x) + \int dx \mu(x) \ln \mu(x) - \int
dx \mu(x) \ln \frac{df^n}{dx}(x) = 0
\end{equation}
In other words, we find that the mean entropy production is zero.
Heuristically, this is quite natural by the very philosophy of the
thermostated dynamics; the change of entropy in the subsystem is
exactly canceled by the change of entropy  in the environment.
That is: the difference in Shannon entropies is given by the
expected phase space contraction. This is known since at least
\cite{And}.

It is true that the above and in particular (\ref{pep}) concerns
the transient regime and that the above calculation cannot be
repeated for the stationary measure as it may be singular. Yet,
this property may be considered as an artifact of the infinitely
fine resolution in $\cal{M}$ and we can remove it by taking a
finite partition $\hat{\cal{M}}$ of $\cal{M}$. We need a
generalization of (\ref{prop}) for dynamics that do not preserve
the phase space volume, $\rho f^{-1} \neq \rho$, with $\rho(dx)=
dx$ the  flat measure on $\cal{M}$. Using the notation of Section
\ref{sec: er}, we now get
\begin{multline}
  \frac{\prob_{\hat\mu\times\rho}
    (x_n\in M_n,\ldots,x_0\in M_0)}
    {\prob_{\hat\mu_n\pi\times\rho}(x_n\in \pi M_0,\ldots,x_0\in \pi M_n)} =
\\
  \frac{\hat\mu(M_0)}{\hat\mu_n(M_n)}\frac{\rho(M_n)}{\rho(M_0)}
    \frac{\rho(\cap_{j=0}^{n} f^{-j} M_j)}{\rho f^n(\cap_{j=0}^{n} f^{-j} M_j)}
\end{multline}
by using again the dynamic reversibility of the map $f$. In the
stationary regime, the formal analogue of the entropy production
rate equals
\begin{multline}
  \lim_{n} \frac{1}{n}
    \ln \frac{\prob_{\hat\mu\times\rho}(x_n\in M_n,\ldots,x_0\in M_0)}
    {\prob_{\hat\mu_n\pi\times\rho}(x_n\in \pi M_0,\ldots,x_0\in \pi M_n)} =
\\
  \lim_{n} \frac{1}{n} \ln \frac{\rho(\cap_{j=0}^{n} f^{-j} M_j)}
    {\rho f^n(\cap_{j=0}^{n} f^{-j} M_j)}
\end{multline}
Note that while this is true for every finite partition
$\hat{\cal{M}}$, it fails for the finest partition where
$\hat{\cal{M}}$ would coincide with the original phase space
$\cal{M}$.
The above formula may be further elaborated, assuming that the
partition $\hat{\cal{M}}$ is generating for $f$. (This would not
be true for the partition $\hat\Gamma$ corresponding to the
physical coarse-graining induced by a set of thermodynamic
variables). Let $x \in \cal{M}$ be fixed and choose $M_j = M(f^jx)$. 
Using the notation $M^{(n)}_{x} = \cap_{j=0}^{n} f^{-j} M_j$,
we have $M^{(n+1)}_x \subset M^{(n)}_x$ and 
$\cap_{n} M^{(n)}_x = \{x\}$. 
Suppose now that the following limits are equal:
\[
  \lim_n \frac{1}{n} \sum_{k=0}^{n}
    \ln \frac{\rho(f^k M^{(n)}_x)}{\rho f(f^k M^{(n)}_x)}
  = \lim_n \frac{1}{n} \sum_{k=0}^{n} \ln \frac{d\rho}{d (\rho f)}(f^k x)
\]
Clearly, $(d \rho f/d \rho) (x)$ is a general form of the phase
space contraction (the Jacobian determinant of $f$).  The
right-hand side takes its ergodic average.  If we sample $x\in
\cal{M}$ from the flat measure $\rho$, we could suppose that these
ergodic averages converge to the expected phase space contraction
for some distribution $\mu$ on $\cal{M}$.  That would for example
be guaranteed under some chaoticity assumptions for the dynamics
$f$; in particular if the dynamical system allows a SRB state
$\mu$, \cite{ru}. We can then combine the previous two relations
and find that, for $\rho$-almost every $x \in \cal{M}$, the mean
entropy production rate gets the form
\begin{equation}\label{mept}
  \lim_{n} \frac{1}{n}
    \ln \frac{\prob_{\hat\mu\times\rho}(x_n\in M(f^n x),\ldots,x_0\in M(x))}
    {\prob_{\hat\mu_n\pi\times\rho\pi}
    (x_n\in \pi M(x),\ldots,x_0\in \pi M(f^n x))}
  = \bbE_{\mu} \Bigl( \ln \frac{d\rho}{d (\rho f)} \Bigr)
\end{equation}
This is exactly the mean entropy production rate one works with in
thermostated dynamics, see e.g. \cite{ru}.  Comparing it with
(\ref{gep}) and (\ref{gepc})-(\ref{gepct}), it does indeed replace
the mean entropy production as computed from the algorithms in
Sections \ref{sec: er} and \ref{sec: cou}.

\section*{Acknowledgment}

We thank S. Goldstein for insisting on clarifying the connections
with the Boltzmann entropy. We thank J. Bricmont for reading a
first draft of the paper. We thank both for useful discussions.

\appendix
\section{Kac ring model}

The scheme of Section \ref{sec: er} can also be applied to every model
dynamics sharing the property of dynamical reversibility with
Hamiltonian dynamics. To illustrate this and in order to specify
some quantities that have appeared above, we briefly discuss the
so called Kac ring model. We refer to the original \cite{K} for
the context and to \cite{bric} for
more discussion.\\
The microscopic state space is $\Omega=\{-1,+1\}^N$.  Its elements
are denoted by $x=(\eta;v)=(\eta_1,\ldots,\eta_N;v)$ and the
kinematic time-reversal is $\pi x = (\eta_1,\ldots,\eta_N;-v)$.
The microscopic dynamics $f_\varepsilon$ depends on parameters
$\varepsilon_i=\pm 1, i=1,\ldots,N$, and is defined as
\begin{align}\nonumber
  f_\epsilon(\eta_1,\ldots,\eta_N;+1) &\equiv
  (\varepsilon_N\eta_N,\varepsilon_1\eta_1,\ldots,\varepsilon_{N-1}\eta_{N-1};+1)
\\\nonumber
  f_\epsilon(\eta_1,\ldots,\eta_N;-1) &\equiv
  (\varepsilon_1\eta_2,\varepsilon_2\eta_3,\ldots,\varepsilon_{N} \eta_{1};-1)
\end{align}
so that $f_\varepsilon = \pi f_\varepsilon^{-1} \pi$ (dynamic
reversibility).  The only information about the parameters is that
$\sum_i\varepsilon_i = mN$ for some fixed $m$.\\
Since the ``velocity'' $v$ is conserved, we can as well study the
dynamics on $\Gamma = \{-1,+1\}^N$ (fixing $v=+1$) and to each
microstate $\eta$ we associate the macroscopic variable
$\alpha(\eta) \equiv \sum_i \eta_i/N$.  This introduces the
partition $\hat\Gamma$ containing $N+1$ elements. For example, the
set $M(\eta)\in \hat\Gamma$ contains all the $\sigma \in \Gamma$
for which $\alpha(\eta)=\alpha(\sigma)$ and $|M(\eta)| =
C_N((\alpha(\eta)+ 1)N/2)$ the binomial factor. Trajectories can
therefore be identified with a sequence of macroscopic values
$\alpha_j$. We are interested in the case of finite (but possibly
long) trajectories while taking $N$ extremely large. In the
simplest approximation, this means that we  let $N\rightarrow
+\infty$.  It can be shown that for the overwhelming majority of
the $\varepsilon_i$, the macroscopic value $\alpha_n$ after $n$
time steps behaves as $\alpha_n = m^n \alpha_0$ with $\alpha_0$
the initial macro-value. The limiting evolution on the level of
macrostates is therefore deterministic but not time-reversal
invariant.  Equilibrium corresponds to $\alpha=0$.  The entropy
production rate (per degree of freedom) when the system has
macro-value $\alpha$ is $(1+\alpha) \ln \sqrt{1+\alpha} +
(1-\alpha) \ln \sqrt{1-\alpha} - (1+ m\alpha) \ln \sqrt{1+m\alpha}
- (1- m \alpha) \ln \sqrt{1- m\alpha} = (1-m^2)\alpha^2/2$ up to
second order in $\alpha$ .

Various examples of applying exactly the algorithm of Section
\ref{sec:markov} to compute the entropy production and to study
its fluctuations have appeared before, see \cite{M,MRV,mm}.  We
just add here how the Markov approximation for the Kac ring model
looks
like.\\
The transition probability can be read from (\ref{markov}): for
finite $N$, $p(\alpha,\alpha')$ is the probability that the
macroscopic value equals $\alpha'$ after one time step, when the
process was started from a randomly chosen $\eta$ with macroscopic
value $\sum_i \eta_i/N = \alpha$:
\begin{equation}\nonumber
  p(\alpha,\alpha') = \frac 1{C_N((\alpha+1)N/2)} |\{\eta\in \Gamma:
  \sum_i\eta_i =\alpha N
\text{ and }
  \sum_i\varepsilon_i \eta_i = \alpha' N\}|
\end{equation}
Depending on the parameters $\varepsilon_i$, this will often be
zero, certainly when $N$ is large and $\alpha'$ is far from equal
to $m\alpha$.  On the other hand, when $\alpha'=m\alpha \pm
\sqrt{1-m^2}/\sqrt{N}$, the transition will be possible but damped
as $\exp[-N(\alpha'-m\alpha)^2/2(1-m^2)]$. It is therefore
interesting to study the evolution on the level of the rescaled
variables $\sqrt{N}\alpha$; these are the fluctuations. This takes
us back to Section \ref{sec: om}.  In equation (\ref{om}), we
should take $R=1/(1-m)$ and $s=1$.  The solution of the Rayleigh
principle is of course here found from maximizing the transition
probability $p(\alpha,\alpha')$ and this happens when
$\alpha'=m\alpha$.  As always with this principle, see \cite{ja1},
this does not teach us anything new; it only gives a variational
characterization of the hydrodynamic evolution.

\section{Hamiltonian dynamics of composed systems}

In order to clear up the content of assumptions A1 and A2 of
Section \ref{sec: cou}, we demonstrate here how it naturally
emerges in the framework of Hamiltonian dynamics. We again have in
mind a composed system consisting of a {\it system} thermally
coupled to an environment, the latter having the form of a few
subsystems (reservoirs). \\
Let $T$ be a finite set whose elements label the individual
particles of the total system. That means that we are really
considering a solid (and not a fluid).  To every particle $i \in
T$, there is associated a position and momentum variable $x_i
\equiv (q_i,p_i) $.  Given a configuration $x$, we put $x_{\La}
\equiv (q_{\La},p_{\La})$ for the coordinates of particles
belonging to the system $\La \subset T$. We thus decompose the set
of particles $T$ by splitting the total system into a {\it system}
and $m$ reservoirs, $T = \La \cup V^1 \cup \ldots \cup V^m$. We
assume that the Hamiltonian of the total system may be written in
the form $H(x) = H^0(x_{\La}) + \sum_{k=1}^{m} H^k(q_{\bar
V^k},p_{V^k})$ where $\bar V^k \equiv V^k \cup \partial_k \La$ and
$\partial_k \La \subset \La$ is the set of all particles of the
system coupled to the $k-$th reservoir. Moreover, we need the
assumption that the reservoirs are {\it mutually separated} in the
sense $\partial_k \La \cap
\partial_\ell \La = \emptyset$ whenever $k \neq \ell$. To be
specific, consider the following form of the Hamiltonian:
\begin{align}
  H^0(x_{\La}) &= \sum_{i\in \La} \bigl[ \frac{p_i^2}{2m_i} + U_i(q_i) \bigr]
    + \sum_{(ij) \subset \La} \Phi(q_i - q_j)
\\
  H^k(q_{\bar V^k},p_{V^k}) &= \sum_{i\in V^k}
    \bigl[ \frac{p_i^2}{2m_i} + U_i(q_i) \bigr]
    + \sum_{(ij) \subset V^k} \Phi(q_i - q_j)
    + \sum_{i \in V^k \atop j \in \partial_k \La} \Phi(q_i - q_j)
\end{align}

For what follows, we consider another decomposition of the energy
of the system in the form $H^0(x_{\La}) = h^0(q_{\La},p_{\La^0}) +
\sum_{k=1}^{n} h^k(x_{\partial_k \La})$ where $\La^0 \equiv \La
\setminus \cup_k \partial_k \La$. We can take, for instance,
\begin{align}
  h^0(q_{\La},p_{\La^0}) &= \sum_{i\in \La^0}
    \bigl[ \frac{p_i^2}{2m_i} + U_i(q_i) \bigr]
    + \sum_{(ij) \subset \La} \Phi(q_i - q_j)
\\
  h^k(x_{\partial_k \La}) &= \sum_{i\in \partial_k \La}
    \bigl[ \frac{p_i^2}{2m_i} + U_i(q_i) \bigr]
\end{align}
If the trajectory $\om(\tau) \equiv (q(\tau),p(\tau))$ is a
solution of the Hamiltonian equations of motion, then the
time-derivative of the energy of each reservoir is in terms of
Poisson brackets:
\begin{equation}
\begin{split}
  \frac{dH^k}{d\tau}(\om(\tau))
    &= \{H^k, H\}(\om(\tau)) = \{H^k, H^0\}(\om(\tau))
\\
    &= \{H^k, h^k\}(\om(\tau))
\end{split}
\end{equation}
A similar calculation yields
\begin{equation}
\begin{split}
  \frac{dh^k}{d\tau}(\om(\tau)) &= \{h^k, H\}(\om(\tau))
\\
  &= \{h^k, H^0\}(\om(\tau)) + \{h^k, H^k\}(\om(\tau))
\end{split}
\end{equation}
where in the last equality we used the assumption that the
reservoirs are mutually separated. Combining the above equations
and integrating them over the time interval $(t_0,t)$ one gets
\begin{multline}\label{plau}
  H^k(\om(t)) - H^k(\om(t_0)) =
    h^k(\om_{\partial_k \La}(t_0)) - h^k(\om_{\partial_k \La}(t))
\\
  - \int_{t_0}^{t} d\tau \sum_{i\in \partial_k \La \atop j\in \La^0}
    \frac{p_i(\tau)}{m_i} \Phi'(q_i(\tau) - q_j(\tau))
\end{multline}
Notice that the right-hand side depends only on the restriction
$\om_{\La}(t)$ of the trajectory $\om$. Therefore, assuming that
$\om(t)$, $\om'(t)$ are two solutions of the equations of motion
such that $\om_{\La}(\tau) = \om'_{\La}(\tau)$, $\tau \in
(t_0,t)$, the heat flow into the $k-$th reservoir,
$Q^k_{(t_0,t)}(\om) \equiv H^k(\om(t)) - H^k(\om(t_0))$ satisfies
$Q^k_{(t_0,t)}(\om) = Q^k_{(t_0,t)}(\om')$. Put differently, the
heat current into each reservoir, being a state quantity from the
point of view of the total system, is also a functional of the
(complete) trajectory of the system. This motivates assumption A2.
Moreover, from the above calculation, also the assumption A1
becomes plausible as we can expect that the right-hand side of
(\ref{plau}) is of order
$(t-t_0)|\partial_k \La|$.\\
{\bf Remark:} Note that the decomposition of the total energy into
local parts is not unique due to the presence of interaction. The
above claim is only true for the decomposition in which the
interaction energy between the system end each reservoir is taken
as part of the energy of the reservoir. However, the difference
between this reservoir energy and others can only be of the order
of $|\partial_k \La|$ which is again sufficient.  Furthermore, all
possible decompositions become undistinguishable in the regime of
weak coupling.

\section{Minimum entropy production principle}\label{sec: MEPP}

In this appendix we examine the validity of the minimum entropy
production principle in case of Markov chains breaking the
detailed balance condition, as promised at the end of Section
\ref{sec:markov}.. We use the same notation as there, namely we
consider a continuous time Markov chain $(X_t)$ on a finite state
space with transition rates $k_{\ve}(M,M')$. The latter are
parameterized by $\ve$ measuring the distance from equilibrium.
More precisely, let $k_{\ve}(M,M') = k_{0}(M,M') \exp[\ve
\psi(M,M') / 2]$ where the Markov chain with rates $k_0(M,M')$ has
a unique reversible measure $\mu_0$, \ie,
\begin{equation}
  \mu_0(M) k_0(M,M') = \mu_0(M') k_0(M',M)
\end{equation}
We also assume that $\mu_0(M) \neq 0$ for all $M$. The stationary
measure $\mu_{\ve}$ is a solution of the equation
\begin{equation}
  \sum_{M} [ \mu_{\ve}(M) k_{\ve}(M,M')
  - \mu_{\ve}(M') k_{\ve}(M',M) ] = 0
\end{equation}
for all $M'$. Write this measure in the form $\mu_{\ve} = \mu_0 (1
+ \ve f + o(\ve))$ with the normalization condition $\sum_{M}
\mu_0(M) f(M) = 0$. Then a simple calculation yields the following
(linearized) equation for stationarity:
\begin{equation}\label{lin}
  \sum_{M} \mu_0(M) k_0(M,M') [ f(M) - f(M') + \psi^{\text{as}}(M,M') ] = 0
\end{equation}
where we used $\psi^{\text{as}}(M,M')$ to denote the asymmetrical
part of the driving, $\psi^{\text{as}}(M,M') = [\psi(M,M') -
\psi(M',M)]/2$. This equation with the constraint $\sum_{M}
\mu_0(M) f(M) = 0$ has always a solution, we will assume it is
unique. Notice that, up to first order in $\ve$, only the
asymmetric part of the driving deforms the stationary measure.

We now compare this result with that of the minimum entropy
production principle. Recall that the entropy production rate is
the functional on measures
\begin{equation}\label{eq: linear}
  \si_{\ve}(\mu) = \sum_{M,M'} \mu(M) k_{\ve}(M,M')
    \ln \frac{\mu(M) k_{\ve}(M,M')}{\mu(M') k_{\ve}(M',M)}
\end{equation}
The first observation is that it is convex. So, the constrained
variational problem $\si_{\ve}(\mu^{\star}) = \min$, $\sum_{M}
\mu^{\star}(M) = 1$, is equivalent to solving the equation
\begin{equation}
  \frac{\de}{\de\mu} \bigl[ \si_{\ve} - \la\sum_M \mu(M) \bigr](\mu^{\star}) = 0
\end{equation}
together with $\sum_{M} \mu^{\star}(M) = 1$. We again linearize
this equation by writing $\mu^{\star}_{\ve} = \mu_0 (1 + \ve
f^{\star} + o(\ve))$ and after some calculation we get
\begin{equation}
  \frac{1}{\mu_0(M)} \sum_{M'} \mu_0(M) k_0(M,M')
  [ f^{\star}(M) - f^{\star}(M') + \psi^{\text{as}}(M,M') - \la ] = 0
\end{equation}
Observe that for $\la = 0$ this equation is equivalent to
\eqref{lin}. Therefore, if the minimizing point $\mu^{\star}$ is
unique, it must correspond to
$f^{\star} \equiv f$ with $f$ being the normalized solution of \eqref{lin}. \\
\\
Note that in higher orders the minimum entropy production
principle fails as a variational principle for the stationary
measure. But even to linear order, outside the context of Markov
processes, the principle can be questioned both for its
correctness and for its usefulness, see \cite{ja1}.

\section{Systems in local thermodynamic equilibrium}\label{sec: LTE}

In this appendix we connect our presentation of Section \ref{sec:
cou} with the standard formulations of irreversible
thermodynamics. We go about this in a rather formal way, trying to
safeguard the
simplicity of the explanation. \\
We consider the system itself to be large (yet small when compared
with the reservoirs) and we split it further into (still large)
subsystems around a spatial point $r$. We assume that these
subsystems are in (local) equilibrium so that the change of
entropy $\hat{S}_B^0(\bar\om_n) - \hat{S}_B^0(\bar\om_0)$ of the
system (appearing in (\ref{cors}) or (\ref{de})) is a change of
(maximal) Boltzmann entropy when the energy in the subsystem
around $r$ moves from the value $U(r,0)$ to $U(r,t)$.  That is,
\[
\hat{S}_B^0(\bar\om_n) - \hat{S}_B^0(\bar\om_0) = \sum_{r}
[S_B(r,U(r,t)) - S_B(r,U(r,0))]
\]
where $S_B(r,U)$ is the logarithm of the phase space volume of the
subsystem around $r$ corresponding to energy value $U$.
  We start from (\ref{de}). It
gets the form
\begin{multline}
  \De S_B(\bar\om) = \sum_{r} [S_B(r,U(r,t)) - S_B(r,U(r,0))]
    + \sum_k \beta_k \De E^k(\bar\om)
\end{multline}
Here the first sum runs over the subsystems of the system under
consideration, while the second sum is taken over all reservoirs.
We introduce the temperature of the $r$-subsystem as 
$\be(r,\tau) = (\partial S_B / \partial U)(r,U(r,\tau))$, and then
\begin{equation}
  \De S_B(\bar\om) = \sum_{r} \int_{0}^{t} d\tau
     \be(r,\tau) \frac{dU}{d\tau}(r,\tau)
    + \sum_{k} \be_k \De E^k(\bar\om)
\end{equation}
We use $J(r,r',\tau)$ to denote the energy current at time $\tau$
from the $r$-subsystem to the $r'$-subsystem. Similarly,
$J^k(r,\tau)$ stands for the energy current from the $r$-subsystem
to the $k-$th reservoir. The conservation of energy then implies
the equalities
\begin{equation}
  \frac{dU}{d\tau}(r,\tau) + \sum_y J(r,r',\tau) + \sum_k
  J^k(r,\tau)=0
\end{equation}
and
\begin{equation}
  \De E^k(\om) = \sum_r \int_0^t d\tau\; J^k(r,\tau)
\end{equation}
The currents are antisymmetric: $J(r,r',\tau) = -J(r',r,\tau)$.
The entropy production now becomes
\begin{equation}
  \De S_B(\bar\om) = \int_{0}^{t} d\tau \Bigl[ \sum_k \sum_r (\be_k - \be(r,\tau))
    J^k(r,\tau) + \sum_{r,r'} \be(r,\tau) J(r',r,\tau) \bigr]
\end{equation}
The first term on the right is a surface sum.  Its origin is the
entropy current. We assume that every subsystem is coupled to at
most one reservoir.  In the continuum, if $r$ is at the boundary
of the system with the $k-$th reservoir, then in fact 
$\be(r,\tau) = \be_k$.  Hence, for the first term, either 
$J^k(r,\tau) = 0$ or $\be(r,\tau) = \be_k$ which makes it vanish. 
When we are dealing with closed systems, 
then $J^k(r,\tau)= 0$ by definition. Using
further the antisymmetry of the bulk currents, we obtain
\begin{equation}
  \De S_B(\bar\om) = \int_{0}^{t} d\tau \sum_{r} \nabla\be(J)(r,\tau)
\end{equation}
where we used the notation
\begin{equation}
  \nabla\be(J)(r,\tau) \equiv
  \sum_{r'} \frac{\be(r',\tau) - \be(r,\tau)}{2} J(r,r',\tau)
\end{equation}
This is already close to the standard formulations in which the
entropy production rate equals a thermodynamic force times a
current. Indeed, assuming that the decomposition of the system
into subsystems has a natural space structure, say as the regular
$\bbZ^d$-lattice, and that the current exchanges take place only
between neighboring subsystems (via their common interface), we
can write $\nabla\be(J)(r,\tau) \simeq \nabla\be(r,\tau) \cdot
\vec J(r,\tau)$ (the derivative taken in the discrete sense). The
(total) entropy production is then $\De S_B(\bar\om) =
\int_{0}^{t} d\tau \sum_r \si(r,\tau)$ with  space-time entropy
production rate
\[
\si(r,\tau) = \nabla\be(r,\tau) \cdot \vec J(r,\tau)
\]
as sought.


\bibliographystyle{plain}

\begin{thebibliography}{10}

\bibitem{And} L. Andrey, {\em The rate of entropy change in
non-Hamiltonian systems}, Phys. Lett. {\bf 11A}, 45-46 (1985).

\bibitem{bric} J. Bricmont, {\em Bayes, Boltzmann and Bohm: Probability in Physics}.
In: Chance in Physics,  Foundations and Perspectives.  Eds. J.
Bricmont, D. D\"urr, M.C. Galavotti, G. Ghirardi, F. Petruccione,
and N. Zanghi, (Springer-Verlag, 2002).


\bibitem{Dorf} J.R. Dorfman, An Introduction to chaos in nonequilibrium statistical
mechanics. (Cambridge University Press, Cambridge, 1999).

\bibitem{ecm}
D.J. Evans, E.G.D. Cohen and G.P. Morriss, {\em Probability of
second law violations in steady flows}, Phys. Rev. Lett. {\bf 71},
2401-2404 (1993).

\bibitem{evans} S. Sarman, D.J. Evans and P.T. Cummings, {\em
Recent developments in non-Newtonian molecular dynamics}, Physics
Reports, Elsevier (Ed. M.J. Klein), {\bf 305}, 1--92 (1998).

\bibitem{ELS}
G.L. Eyink, J.L. Lebowitz and H. Spohn, {\em  Microscopic origin
of hydrodynamic behavior: Entropy production and the steady
state}, Chaos (Soviet-American Perspectives on Nonlinear Science),
ed. D. Campbell, 367--391 (1990).


\bibitem{GC}
G. Gallavotti and E.G.D. Cohen, {\em Dynamical ensembles in
nonequilibrium \mbox{Statistical} Mechanics}, Phys. Rev. Letters
{\bf 74}, 2694--2697 (1995). {\em Dynamical ensembles in
stationary states}, J. Stat. Phys. {\bf 80}, 931--970 (1995).

\bibitem{Gold} S. Goldstein, {\em Boltzmann's Approach to Statistical
Mechanics}. In: Chance in Physics,  Foundations and Perspectives.
Eds. J. Bricmont, D. D\"urr, M.C. Galavotti, G. Ghirardi, F.
Petruccione, and N. Zanghi, (Springer-Verlag, 2002).

\bibitem{jar}
C. Jarzynski, {\em Nonequilibrium Equality for Free Energy
Differences}, Phys. Rev. Lett. {\bf 78}, 2690--2693 (1997).


\bibitem{j} E.T. Jaynes, {\em Gibbs vs Boltzmann Entropies}, Am. J. Phys. {\bf 33},
391--398 (1965). {\em Papers on Probability, Statistics and
Statistical Physics}, Ed. R. D. Rosencrantz (Reidel, Dordrecht
1983).

\bibitem{ja1} E.T. Jaynes, {\em The minimum entropy production
principle}, Ann Rev. Phys. Chem. {\bf 31}, 579--600 (1980). {\em
Papers on Probability, Statistics and Statistical Physics}, Ed. R.
D. Rosencrantz (Reidel, Dordrecht 1983).  Note however that
historically, the Rayleigh principle (Lord Rayleigh, Phil Mag.
{\bf 26}, 776 (1913)), was much closer to the variational
principle in mechanics.

\bibitem{K} M. Kac, {\em Probability and Related Topics in the
Physical Sciences}, (Interscience Pub., New York 1959).

\bibitem{le} J.L. Lebowitz, {\em Reduced Description in
Nonequilibrium Statistical Mechanics}. In: Proceedings of
International Conference on Nonlinear Dynamics, December 1979,
Annals New York Academy of Sciences {\bf 357}, 150--156 (1980).


\bibitem{Lebo} J.L. Lebowitz, {\em Microscopic Origins of Irreversible
Macroscopic Behavior}, Physica A {\bf 263}, 516--527 (1999). Round
Table on Irreversibility at STATPHYS20, Paris, July 22, 1998.


\bibitem{Liu} M. Liu, {\em The Onsager Symmetry Relation and the
Time Inversion Invariance of the Entropy Production}, Archive {\tt
cond-mat/9806318}.

\bibitem{M}
C. Maes, {\em Fluctuation theorem as a Gibbs property\/}, J. Stat.
Phys. {\bf 95}, 367--392 (1999).

\bibitem{MRV}
C. Maes, F. Redig and A. Van Moffaert, {\em On the definition of
entropy production via examples}, J. Math. Phys. {\bf 41},
1528--1554 (2000).


\bibitem{mm}
C. Maes, F. Redig and M. Verschuere, {\em From Global to Local
Fluctuation Theorems}, Moscow Mathematical Journal {\bf 1},
421--438 (2001).


\bibitem{om} L. Onsager and S. Machlup, {\em Fluctuations and Irreversible Processes},
Phys. Rev. {\bf 91}, 1505--1512 (1953).  More generally
Onsager-Machlup consider in their second paper (starting on page
1512) a second-order process for
  extensive variables
$(\alpha,\beta)\equiv (\alpha_1,\ldots,
\alpha_m;\beta_1,\ldots,\beta_m)$ parametrizing the partition
$\hat\Gamma$ that we had before. The difference between the
$\alpha$'s and the $\beta$'s arises from their different behavior
under applying the time reversal $\pi$; the $\beta$'s are the so
called `velocity' variables that change their sign if the time is
reversed; the $\alpha$'s are even functions under time reversal.
In all cases, they are thermodynamic variables that are sums of a
large number of molecular variables.

\bibitem{ru}
D. Ruelle, {\em Smooth Dynamics and New Theoretical Ideas in
Nonequilibrium Statistical Mechanics}, J. Stat. Phys. {\bf 95},
393--468 (1999).

\bibitem{Schn}
J. Schnakenberg, {\em Network theory of behavior of master
equation systems}, Rev. Mod. Phys. {\bf 48}, 571-585 (1976).

\bibitem{Sinai} Ya. G. Sinai, Introduction to Ergodic Theory,
Princeton University Press (1976).

\bibitem{Z} R. Zwanzig, {\em Memory Effects in Irreversible
Thermodynamics}, Phys. Rev. {\bf 124}, 983--992 (1961).  See also:
S. Nakajima, Prog. Theor. Phys. {\bf 20}, 948 (1958); M. Mori,
Prog. Theor. Phys. {\bf 32}, 423 (1965).


\end{thebibliography}

\end{document}